\documentclass[12pt]{article}
\usepackage{a4wide,epsfig,psfrag,amsmath,amssymb,cite,scalefnt}
\usepackage{color}
\usepackage{amsmath,comment,braket}

\graphicspath{ {figs/} }

\newcommand{\be}{\begin{equation}}
\newcommand{\ee}{\end{equation}}
\newcommand{\bea}{\begin{eqnarray}}
\newcommand{\eea}{\end{eqnarray}}

\allowdisplaybreaks

\newcommand{\gsim}{\;\rlap{\lower 3.5 pt \hbox{$\mathchar \sim$}} \raise 1pt
 \hbox {$>$}\;}
\newcommand{\lsim}{\;\rlap{\lower 3.5 pt \hbox{$\mathchar \sim$}} \raise 1pt
 \hbox {$<$}\;}

\begin{document}

\title{\vskip-3cm{\baselineskip14pt
    \begin{flushleft}
      \normalsize TTP18-009
  \end{flushleft}}
  \vskip1.5cm
  Double-Higgs boson production in the high-energy limit: planar master integrals
}

\author{
  Joshua Davies$^{a}$,
  Go Mishima$^{a,b}$,
  Matthias Steinhauser$^{a}$,
  David Wellmann$^{a}$
  \\[1mm]
  {\small\it $^a$Institut f{\"u}r Theoretische Teilchenphysik}\\
  {\small\it Karlsruhe Institute of Technology (KIT)}\\
  {\small\it 76128 Karlsruhe, Germany}
  \\[1mm]
  {\small\it $^b$Institut f{\"u}r Kernphysik}\\
  {\small\it Karlsruhe Institute of Technology (KIT)}
  \\
  {\small\it 76344 Eggenstein-Leopoldshafen, Germany}
}
  
\date{}

\maketitle

\thispagestyle{empty}

\begin{abstract}

  We consider the virtual corrections to the process $gg\to HH$ at NLO
  in the high energy limit and compute the corresponding planar master
  integrals in an expansion for small top quark mass. We provide details
  on the evaluation of the boundary conditions and present analytic results
  expressed in terms of harmonic polylogarithms.

%

\end{abstract}

\thispagestyle{empty}

\sloppy


\newpage


\section{Introduction}

One of the main aims of particle physics in the coming years is the
exploration of the scalar sector of the theory which describes fundamental
interactions, be it the Standard Model or an extension.  One has to clarify
whether the Higgs boson is a fundamental particle and how the
particles of the theory obtain their mass.  A process which helps to find
answers to these questions is the production of Higgs boson pairs, since it is
the simplest process which is sensitive to the triple-Higgs boson
coupling. Although experimentally quite challenging, there is a chance that
double Higgs production will be observed after the the high-luminosity upgrade
of the CERN LHC.

The leading order (LO) corrections to Higgs boson pair production have been
computed in Refs.~\cite{Glover:1987nx,Plehn:1996wb} including the exact
dependence on the top quark mass and the Mandelstam variables.  At
next-to-leading order (NLO), QCD corrections were computed for the first
time in Ref.~\cite{Dawson:1998py} in the infinite top quark mass limit using
an effective theory and an independent cross check was provided
in~\cite{Grigo:2013rya} by performing an asymptotic expansion in the full
theory.  In this way a quantitative estimate of the quark mass
effects could be provided.  Virtual NLO corrections in the large-$m_t$ limit
have also been computed in Ref.~\cite{Degrassi:2016vss}, confirming the results
of Ref.~\cite{Grigo:2013rya}.
Finite top quark mass effects have also been
considered in Ref.~\cite{Maltoni:2014eza}, in which the exact real radiation
contribution is combined with the effective-theory virtual corrections.
Within the effective theory also next-to-next-to-leading (NNLO) contributions
are available~\cite{deFlorian:2013uza,deFlorian:2013jea}. The NNLO result
was completed in Ref.~\cite{Grigo:2014jma} in which the three-loop matching
coefficient of the effective operator for two Higgs bosons and two, three or
four gluons was computed. Note that it differs from that of single
Higgs boson production. The result of~\cite{Grigo:2014jma} has been
complemented by power-suppressed terms in the top quark mass in
Ref.~\cite{Grigo:2015dia}, where the soft-virtual approximation was
constructed.
The resummation of threshold-enhanced logarithms to
next-to-next-to-leading logarithmic (NNLL) accuracy has been performed in
Refs.~\cite{Shao:2013bz,deFlorian:2015moa} and transverse momentum resummation
has been considered in Ref.~\cite{Ferrera:2016prr}.  Differential
distributions through NNLO for various observables were computed in
Ref.~\cite{deFlorian:2016uhr} in the heavy-top limit.  Finally, exact NLO
results became available in Refs.~\cite{Borowka:2016ypz,Borowka:2016ehy} using
a numerical approach for the computation of the two-loop virtual corrections.
More recently these results have been matched to parton showers in
Ref.~\cite{Heinrich:2017kxx}.

In this paper we study a class of massive two-loop four-point
functions with massless external particles. We describe in detail the methods
used for the computation of the amplitudes and in particular the evaluation of
the master integrals.  We aim to study double Higgs boson production
via the process $gg\to HH$. Numerical NLO results are
available~\cite{Borowka:2016ypz,Borowka:2016ehy}, however the calculation of
cross sections is computationally expensive and we want to provide an independent cross check
in the high-energy region. We wish to provide results in terms of compact
analytic expressions which can be used to construct simple
approximations or can be used directly in the kinematic region in which
they are valid. In this paper we provide the first step towards this goal by
considering the part of the amplitude which is expressed in terms of planar master
integrals.

We perform our calculation in the limit of vanishing Higgs boson mass which
provides, as we will demonstrate in Section~\ref{sec::gg2hh1l}, a good approximation
to the general case where $m_H\not=0$. Furthermore, finite Higgs-mass effects can be
incorporated by a simple Taylor expansion.  Recently the amplitudes for
single-Higgs boson plus jet production have been considered in the limit of large Higgs
transverse momentum~\cite{Kudashkin:2017skd}. In this reference an expansion
for small Higgs boson mass has been performed and thus the underlying
integrals are the same as those of our calculation, so part of our findings
can be cross checked against Ref.~\cite{Kudashkin:2017skd}.

In the recent literature one can find several calculations where two-loop
box integrals are also involved. However, the underlying integral families and/or
the kinematics of the external and internal masses are different. For example,
in Ref.~\cite{Bonciani:2016qxi} the amplitude of a Higgs boson and three
partons has been considered.  In the limit $m_H\to 0$ their integrals are 
also the
same as ours. However, this limit cannot been taken since the calculation is
performed in the Euclidean region with the assumption $m_H^{2}<s<0$ and the results are
expressed in terms of multiple polylogarithms, which can not easily be analytically
continued into other regions. Similar arguments apply to other recent calculations
such as~\cite{Becchetti:2017abb} or~\cite{Mastrolia:2017pfy}; analytic results
have been obtained in terms of multiple polylogarithms which can in principle be
evaluated numerically, but are very unwieldy.
This is a another reason why we
have decided to perform an expansion in the high energy limit. Our final
results have a simple structure in terms of harmonic polylogarithms and can be
evaluated numerically in a fast and reliable manner.

An interesting approach to obtain simple and easy-to-evaluate expressions for
$gg\to HH$ at NLO has been developed in Ref.~\cite{Grober:2017uho} where the
large top mass expansion has been combined with expansion terms obtained for
the top threshold. A good approximation of the exact (purely numerical)
result~\cite{Borowka:2016ypz,Borowka:2016ehy} has been constructed by
combining the different kinematic regions using Pad\'e
approximants. Further improvement is expected after incorporating
information about the $gg\to HH$ amplitude at high energies which is the main purpose of
this work.

The remainder of the paper is organized as follows: we introduce our notation
in Section~\ref{sec::notation}. In Section~\ref{sec::gg2hh1l} we briefly consider the
one-loop corrections to $gg\to HH$ in the high-energy limit to provide motivation
for our calculation, and Section~\ref{sec::red} describes the reduction of the
amplitude to master integrals. The main part of the paper is
Section~\ref{sec::master} in which we discuss the calculation of the master
integrals. We describe in detail the method we used to compute the boundary
values necessary for the solution of differential equations for the master integrals.
In this paper we refrain from presenting long formulae, which instead can be
found in the ancillary file of this paper~\cite{progdata}.


\section{\label{sec::notation}$gg\to HH$ amplitude and kinematics}

The amplitude $g(q_1)g(q_2)\to H(q_3)H(q_4)$, where all momenta are incoming,
is conveniently described in terms of the following variables
\begin{eqnarray}
  s=(q_1+q_2)^2\,,\qquad t=(q_1+q_3)^2\,,\qquad u=(q_2+q_3)^2\,,
\end{eqnarray}
where we have that
\begin{eqnarray}
  q_1^2=q_2^2=0\,,\qquad  q_3^2=q_4^2=m_H^{2}\,,\qquad s+t+u=2m_H^{2}\,.
\end{eqnarray}
Here we make the approximation $m_H=0$ which significantly
simplifies the two-loop calculation. Then we have instead that
\begin{eqnarray}
  s=2q_1\cdot q_2\,,\qquad t=2q_1\cdot q_3\,,\qquad u=2q_2\cdot q_3=-s-t,
\end{eqnarray}
and integrals will depend on the variables $s,t,m_t^{2}$.  Note that finite
Higgs mass effects can be implemented by a simple Taylor expansion.  Each
integral is proportional to $s^{a_1}(s/\mu^2)^{-\epsilon a_2}$ where
$a_2=1(2)$ at one-(two-)loop order, $a_1$ is its overall mass dimension,
$\mu$ is the renormalization scale and we work in $d=4-2\epsilon$
dimensions.  In our calculation of the master integrals we expand the
integrals for small top quark mass. Thus, effectively we assume that $m_t^{2}
\ll s,t$.

Due to Lorentz and gauge invariance there are only two independent Lorentz
structures and we can write
\begin{eqnarray}
  {\cal M} &=& 
  \varepsilon_{1,\mu}\varepsilon_{2,\nu}
  {\cal M}^{\mu\nu}
  \,\,=\,\,
  \varepsilon_{1,\mu}\varepsilon_{2,\nu}
  \left( {\cal M}_1 A_1^{\mu\nu} + {\cal M}_2 A_2^{\mu\nu} \right)
  \,,
\end{eqnarray}
where
\begin{eqnarray}
  A_1^{\mu\nu} &=& g^{\mu\nu} - \frac{1}{q_{12}}q_1^\nu q_1^\mu\,,\nonumber\\
  A_2^{\mu\nu} &=& g^{\mu\nu} 
  + \frac{ q_{33} }{ q_T^2 q_{12} } q_1^\nu q_2^\mu    
  - \frac{ 2q_{23} }{ q_T^2 q_{12} } q_1^\nu q_3^\mu    
  - \frac{ 2q_{13} }{ q_T^2 q_{12} } q_3^\nu q_2^\mu    
  + \frac{ 2 }{ q_T^2 } q_3^\mu q_3^\nu    
  \,,
\end{eqnarray}
and
\begin{eqnarray}
  q_{ij} &=& q_i\cdot q_j\,,\qquad
  q_T^{\:2} \:\:\:=\:\:\: \frac{2q_{13}q_{23}}{q_{12}}-q_{33}
  \,.
\end{eqnarray}
The projectors to obtain ${\cal M}_1$ and ${\cal M}_2$ from ${\cal M^{\mu\nu}}$
via the relation
\begin{eqnarray}
  {\cal M}_i &=& P_{i,\mu\nu} {\cal M}^{\mu\nu}
\end{eqnarray}
are given by (see also, e.g., Ref.~\cite{Borowka:2016ypz})
\begin{eqnarray}
 P_{1,\mu\nu} &=&
     - \frac{q_{1,\nu} q_{2,\mu} q_{33}}{4 q_{12} q_T^{\:2}}
     - \frac{q_{1,\nu} q_{2,\mu}}{4 q_{12}}
     + \frac{q_{1,\nu} q_{3,\mu} q_{23}}{2 q_{12} q_T^{\:2}}
     + \frac{q_{2,\mu} q_{3,\nu} q_{13}}{2 q_{12} q_T^{\:2}}
     - \frac{q_{3,\mu} q_{3,\nu}}{2 q_T^{\:2}}
\nonumber\\[-1mm]&&
     + \frac{1}{(2-4\epsilon)} \Bigg[
          \frac{q_{1,\nu} q_{2,\mu} q_{33}}{2 q_{12} q_T^{\:2}}
        - \frac{q_{1,\nu} q_{2,\mu}}{2 q_{12}}
        - \frac{q_{1,\nu} q_{3,\mu} q_{23}}{q_{12} q_T^{\:2}}
        - \frac{q_{2,\mu} q_{3,\nu} q_{13}}{q_{12} q_T^{\:2}}
        + \frac{q_{3,\mu} q_{3,\nu}}{q_T^{\:2}}
        + g_{\mu\nu}
     \Bigg]\,,
\nonumber\\[2mm]
 P_{2,\mu\nu} &=&
     \frac{q_{1,\nu} q_{2,\mu} q_{33}}{4 q_{12} q_T^{\:2}}
     + \frac{q_{1,\nu} q_{2,\mu}}{4 q_{12}}
     - \frac{q_{1,\nu} q_{3,\mu} q_{23}}{2 q_{12} q_T^{\:2}}
     - \frac{q_{2,\mu} q_{3,\nu} q_{13}}{2 q_{12} q_T^{\:2}}
     + \frac{q_{3,\mu} q_{3,\nu}}{2 q_T^{\:2}}
\nonumber\\[-1mm]&&
     + \frac{1}{(2-4\epsilon)} \Bigg[
          \frac{q_{1,\nu} q_{2,\mu} q_{33}}{2 q_{12} q_T^{\:2}}
        - \frac{q_{1,\nu} q_{2,\mu}}{2 q_{12}}
        - \frac{q_{1,\nu} q_{3,\mu} q_{23}}{q_{12} q_T^{\:2}}
        - \frac{q_{2,\mu} q_{3,\nu} q_{13}}{q_{12} q_T^{\:2}}
        + \frac{q_{3,\mu} q_{3,\nu}}{q_T^{\:2}}
        + g_{\mu\nu}
     \Bigg]\,.
     \nonumber\\
\end{eqnarray}
The partonic cross section is obtained from $|{\cal M}|^2$ after integration
over the phase space and multiplication by the flux factor.


\section{\label{sec::gg2hh1l}One-loop considerations}

Before providing details on our two-loop calculation we want to investigate
the quality of the high-energy expansion at one-loop order.
In the following we consider the
differential partonic cross section
\begin{eqnarray}
  \frac{ {\rm d}\sigma }{ {\rm d} {{\theta}} }(s,t)
\end{eqnarray}
where the scattering angle ${{\theta}}$ of the Higgs boson in
the center-of-mass frame enters via the following relation,
\begin{eqnarray}                                                                                       
  t &=& -\frac{s}{2}\left(1-\cos{{\theta}}\right)
  \,.
\end{eqnarray}

\begin{figure}[t]
  \centering
  \includegraphics[width=\textwidth]{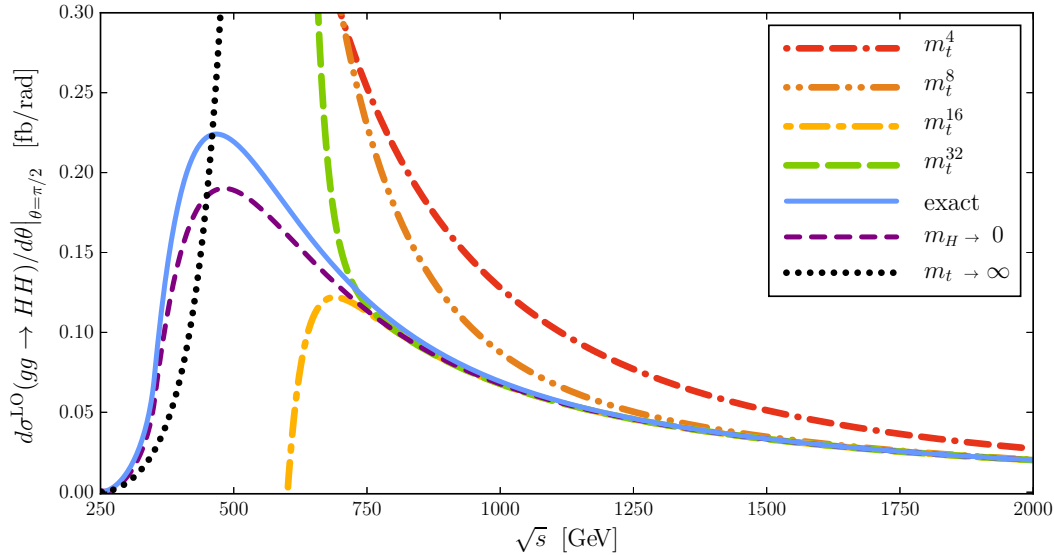}
  \caption{\label{fig::ds_s}${ {\rm d}\sigma }/{ {\rm d} {{\theta}} }$ as a
    function of $\sqrt{s}$ for fixed ${{\theta}} =\pi/2$.}
\end{figure}

In Fig.~\ref{fig::ds_s} we study the $\sqrt{s}$-dependence of ${ {\rm d}\sigma
}/{ {\rm d} {{\theta}} }$ for fixed scattering angle ${{\theta}}$ of 90~degrees.  The
exact result (see the solid curve for $m_H\not=0$ and the short-dashed curve for $m_H=0$) is
compared to various approximation, computed for $m_H=0$, incorporating
high-energy expansions up to $m_t^{32}$ (see the long-dashed and dash-dotted curves).
For comparison we also show the result based on an effective-theory
calculation in which the limit of infinite top quark mass is assumed (dotted
curve).  We observe that, as expected, the high-energy approximations lead to
good results for large values of $\sqrt{s}$. A systematic improvement is
obtained after including higher order expansion terms. For example, for
$\sqrt{s} \approx 1500$~GeV the curves including $m_t^8$ and $m_t^{16}$ terms
agree with each other and the exact (short-dashed) curve which suggests that
an approximation incorporating $m_t^8$ terms works well above this energy.
With the same argument we conclude that $m_t^{16}$ terms are sufficient to describe the exact result
down to $\sqrt{s} \approx 750$~GeV. It seems that one cannot go significantly
below this energy since both the $m_t^{16}$ and the $m_t^{32}$ curves
start to diverge just below $\sqrt{s} \approx 750$~GeV.

In the evaluation of the master integrals we assume $m_t^{2} \ll s,t$.  This
means that the expansion breaks down for ${{\theta}} \to 0,\pi$ (where $|t|$
becomes small) and thus a restricted phase space has to be considered when
performing the integration over ${{\theta}}$. In practical applications 
this does not constitute a big problem since ${{\theta}} \to 0,\pi$ corresponds
to the forward and backward scattering of the Higgs boson, where no measurement
can be performed.
Furthermore, the bulk of the cross section is provided by the central region.
For example, if we restrict $0.25\pi <{{\theta}} < 0.75 \pi$ in the exact one-loop
corrections we cover around 70\% of the full cross section for $\sqrt{s}=1000$~GeV.

Apart from providing an independent and analytically simple expression in the
high energy region, which can be used as a cross-check of the exact (numerical)
results, our expressions also serve as input of the method based on Pad\'e
approximants~\cite{Grober:2017uho} as already mentioned in the Introduction.


\section{\label{sec::red}Reduction to master integrals}

We generate our amplitudes with the program {\tt qgraf}~\cite{Nogueira:1991ex}
and use {\tt q2e} and {\tt exp}~\cite{Harlander:1997zb,Seidensticker:1999bb}
to rewrite the output to {\tt FORM}~\cite{Ruijl:2017dtg} notation. {\tt exp}
is also used to assign to each Feynman diagram an integral family which is
defined according to the topology and mass distribution of the internal lines.
For our application we have defined 34 families.  We use {\tt FORM} to express
the amplitude for each diagram as a linear combination of scalar integrals of
a given family.

We use the {\tt C++} version of {\tt FIRE}~\cite{Smirnov:2014hma} for the
reduction of all scalar integrals from each family to master integrals,
with {\tt LiteRed}~\cite{Lee:2012cn,Lee:2013mka} providing {\tt \#rules} for
{\tt FIRE}.  All families were reduced using the publicly available version
{\tt FIRE 5.2}.  Some families were also reduced using the development
versions {\tt FIRE 5.5, 5.6} (which use more information from the {\tt
  LiteRed} files) in order to check whether the number of master integrals
produced was smaller. This was the case, but the number of master integrals
was still not minimal. We describe our procedure to obtain a minimal set of
master integrals below. The {\tt Mathematica}-readable tables generated by
{\tt FIRE} are transformed to {\tt FORM} {\tt Fill} statements, so that the
reduction can be applied to the amplitude in {\tt FORM} using a {\tt
  TableBase}. The reduction rules are heavily manipulated in {\tt FORM} before
creating this {\tt TableBase}.

After the {\tt FIRE} reduction, each integral family contains between 7 and 77
master integrals (1395 in total (1+2 loops)). One must minimize the number of
master integrals between all families; the use of {\tt FIRE}'s {\tt
  FindRules[]} command yields a total of $231 = 10+221$ master
integrals. This does not constitute a minimal set. We use the following
procedure, implemented in {\tt FORM}, to find and eliminate ``master
integrals'' which are in fact a linear combination of other integrals of the
set.

\begin{enumerate}
\item For each $n$-line master integral with a dot on one of its lines,
  generate the $n-1$ integrals which instead have a dot on one of the other lines. Append these
  integrals to the set of integrals from the amplitude.
  Ensure that each integral in this extended set is present in the reduction tables.
\item Apply relations from {\tt FindRules[]} to the extended set of integrals and
  consider equations of the form:
  \begin{equation}
  	\label{eq::findrulesrelations}
  	\mbox{\texttt{FindRules[ I ] == I}}.
  \end{equation}
  Apply the reduction tables to these equations and discard all trivial
  equations. One obtains a set of non-trivial equations which relate some integrals
  in the original set of master integrals.
\item Use these equations to construct reduction relations into a final linearly
  independent set of master integrals. We solve the equations to obtain the ``most
  complicated'' (highest line count) integrals in terms of simpler integrals.
\end{enumerate}

In Step 2, we consider such equations for all integrals $\texttt{I}$ of
complexity $<9$, where we define the complexity as the sum of the absolute
values of the propagator powers. The reduction rules for higher complexity
integrals contain coefficients which are too large to efficiently manipulate
with {\tt FORM}'s {\tt PolyRatFun}. Despite this, the set of equations
contains many redundancies. That these equations are all satisfied increases
our confidence that our final set of master integrals is a minimal
set. Additionally, {\tt FindRules[]} maps many integrals into different
integral families, so this procedure shows some consistency between our
families.  We note that no approximation (except $m_H=0$) is applied during
the reduction procedure. In particular, we retain the exact $m_t$ dependence.

Following this procedure we reduce the number of two-loop master integrals
from 221 to 161.  A list of them, and all 10 one-loop\footnote{Note that only
  7 of these 10 master integrals appear in the amplitude.}  master integrals,
can be found in Appendix~\ref{app::MIs}.  At one-loop order we obtain the
minimal set of ten master integrals simply by applying {\tt FindRules[]} to
the master integrals of the three one-loop families.  The additional
two-loop reduction relations are applied to the {\tt FIRE} reduction relations
before we create the {\tt FORM TableBase} which we apply to the amplitude and
to the right-hand-side of differential equations (see
Subsection~\ref{sub::diffeq}).

The computation of these master integrals is described in Section~\ref{sec::master}.


\section{\label{sec::master}Calculation of master integrals}

For the calculation of the master integrals we use the method of differential
equations~\cite{Kotikov:1990kg,Gehrmann:1999as}. We solve the differential 
equations using an appropriate ansatz which is described in
Subsection~\ref{sub::diffeq}. The boundary conditions (see Subsection~\ref{sub::BC})
are fixed by evaluating the master integrals in the limit
$m_t\to 0$.  In some cases it is sufficient to evaluate the integrals in this limit for
fixed $t=s=-1$.

\subsection{\label{sub::diffeq}Differential equations}

We compute the master integrals in an unphysical
region where two Mandelstam variables ($s$ and $t$) are negative and $u$ is positive.
In this region, the integrals which we compute are real valued.
We can analytically continue results obtained here into the physical region.

For each master integral we have three differential equations which are
obtained by taking derivatives w.r.t. $m_t^{2}$, $s$ and $t$.  The
derivatives are computed using {\tt LiteRed}.  Note that only
two of the three differential equations are needed to construct the
result. The third provides a consistency check. The generation of the system of
differential equations requires the extension of the {\tt FIRE} reduction tables. Note,
however, that the additional integrals which are required are not difficult to reduce.

Differentiating the vector of master integrals, $(MI)$, w.r.t. $x=t,m_t^{2}$ and applying
the reduction tables to the result leads to systems of equations
\begin{eqnarray}
  \frac{{\rm d}}{{\rm d}x} (MI) &=& K_x \cdot (MI)
  \,,
  \label{eq::diffeq}
\end{eqnarray}
where $K_x$ is a square matrix.

To solve the differential equations we follow two approaches.
In the first, we 
make an ansatz for each master integral which is suitable to describe the solution
in the limit $m_t\to 0$. Guided by asymptotic expansion
we use (see also Refs.~\cite{Melnikov:2017pgf,Kudashkin:2017skd})
\begin{equation}
  I = 
  \sum_{n_1= n_1^{\rm min} }^\infty 
  \sum_{n_2= n_2^{\rm min} }^\infty 
  \sum_{n_3=0}^{ 2l+n_1 }
  c(I,n_1,n_2,n_3,s,t)\,
  \epsilon^{n_1} \left(m_t^{2}\right)^{n_2} \left(\log(m_t^{2})\right)^{n_3}
  \,,
  \label{eq::ansatz}
\end{equation}
where $l$ is the number of loops.
To determine the coefficients $c$ of the ansatz
we use the following procedure:
\begin{itemize}
\item[(1)] Use the differential equation for $t$ and determine the coefficients of
the leading terms in the $m_t\to 0$ expansion.  This requires the solution of a
system of first-order differential equations for the $t$-dependent
coefficients $c$. Boundary conditions are needed for one specific value of $t$,
e.g., for $t=s=-1$.

\item[(2)] Use the differential equation for $m_t^{2}$ to obtain relations between the
coefficients of the higher order $m_t^{2}$ terms and the leading terms
determined in (1).
Since the $m_t$ dependence is explicit in the ansatz one only has to solve a system
of linear equations.

\item[(3)] The results for the master integrals are inserted into the differential
equation for $s$, which must be satisfied.
\end{itemize}

At one-loop order the matrix $K_t$ in Eq.~(\ref{eq::diffeq}) 
has a triangular structure. Thus, starting from the simplest integral, one can
run through the vector of master integrals and solve the system integral-by-integral.

At two-loop order $K_t$ in Eq.~(\ref{eq::diffeq}) rather has a block-triangular
structure. It contains blocks of up to four integrals, which
form coupled systems of differential equations. The integrals within these blocks
must be determined together.

Using the above approach based on $t$-independent boundary conditions we
were not able to obtain results for all two-loop master integrals.
Presumably this is due to an inconvenient choice of our set of master integrals.
For this reason we
developed a second approach where we determine the master integrals in the
limit $m_t^{2}\to0$, keeping the full $t$ dependence.  Afterwards we only have to
solve the $m_t^{2}$ differential equation which, as mentioned above, reduces
to solving a system of linear equations. This approach provides
results for all master integrals. Where possible, we compared the results
of the two approaches and found complete agreement.

Let us end this subsection by making a brief comment on the possibility to
introduce a canonical basis~\cite{Henn:2013pwa} for our master integrals.
We made several attempts to produce such a basis using the publicly available programs
{\tt Fuchsia}~\cite{Gituliar:2017vzm} and {\tt CANONICA}~\cite{Meyer:2017joq}.
We were not able to obtain a canonical basis for our master integrals for all sectors.
Since we are interested in the small-$m_t$ limit
we did not insist on obtaining a canonical basis.

\subsection{\label{sub::BC}Boundary conditions}

The main tools which we use to compute the boundary conditions are the
method of regions~\cite{Beneke:1997zp,Smirnov:2012gma} and Mellin-Barnes
techniques (see, e.g.,~\cite{Smirnov:2012gma}).  Additionally, we make use of
the PSLQ algorithm~\cite{PSLQ} and exploit the anticipated dependence on
irrational numbers of our final result to obtain exact expressions.

In the following we provide details of each
step of the calculation and give concrete examples for the master integral
$G_{6}(1,1,1,1,1,1,1,0,0)$. See Appendix~\ref{app::MIs} for definitions of
the integral families.

In this section we assume the scaling
\begin{align}
m_t^{2}\sim \chi,\quad
s\sim 1,\quad
t\sim 1,
\end{align}
where the parameter $\chi\ll 1$ is introduced for convenience.

To begin, we express the Feynman integral in its alpha representation
(using the routines provided in {\tt FIESTA}~\cite{Smirnov:2015mct}). This is
a convenient starting point to apply the method of regions.  For example,
the integral $G_{6}(1,1,1,1,1,1,1,0,0)$ is expressed as
\begin{eqnarray}
  J &=&
  {(\mu ^2)^{2\epsilon}}
  {e^{2\epsilon\gamma_E}}
  \left( \prod_{i=1}^7 \int_0^\infty \alpha_i^{\delta_i} \right)
  \mathcal{U}^{-d/2}
  \exp \left( -\mathcal{F}/\mathcal{U} \right)
  \,,
  \label{eq:alpha-rep.original}
\end{eqnarray}
where the functions $\mathcal{U}$ and $\mathcal{F}$,
the so-called first and second Symanzik polynomials,
are given by
\begin{eqnarray}
 \mathcal{U}&=&
  \alpha_1 \alpha_4+\alpha_1 \alpha_5+\alpha_1 \alpha_6+\alpha_1
  \alpha_7+\alpha_2 \alpha_4+\alpha_2 \alpha_5+\alpha_2 \alpha_6
  +\alpha_2 \alpha_7
  \nonumber\\&&\mbox{}
  +\alpha_3 \alpha_4+\alpha_3 \alpha_5+\alpha_3
  \alpha_6+\alpha_3 \alpha_7+\alpha_4 \alpha_7+\alpha_5 \alpha_7
  +\alpha_6 \alpha_7
  \,,
  \nonumber\\
  \mathcal{F}&=&
  m_t^{2} \left(
    \alpha_1+\alpha_2+\alpha_3+\alpha_4+\alpha_5+\alpha_6\right)  \mathcal{U}
    -t \left(\alpha_2 \alpha_5 \alpha_7\right)
     -s\left( \alpha_1 \alpha_3\alpha_4+\alpha_1 \alpha_3 \alpha_5
     \right. 
  \nonumber\\&&\mbox{}
    \left.
       +\alpha_1\alpha_3 \alpha_6 
      +\alpha_1 \alpha_3 \alpha_7+\alpha_1 \alpha_4 \alpha_6+\alpha_1
      \alpha_4 \alpha_7+\alpha_2 \alpha_4 \alpha_6
   +\alpha_3 \alpha_4 \alpha_6
  +\alpha_3 \alpha_6 \alpha_7+\alpha_4 \alpha_6 \alpha_7
  \right)  \,.
  \nonumber\\
  \label{eq::UF}
\end{eqnarray}
In Eq.~\eqref{eq:alpha-rep.original} we have introduced analytic regularization parameters
$\delta_i$ to regularize collinear divergences, which appear later in the
calculation.  The original integral is obtained by taking the
sequence limit $\delta_i\to 0$ for all $\delta_i$.

To implement the asymptotic expansion for $\chi \to0$ we use the program {\tt
  asy.m}~\cite{Pak:2010pt}. It provides scaling rules for the alpha parameters for
the various regions which have to be considered.  For the integral in
Eq.~(\ref{eq:alpha-rep.original}) there are 13 relevant regions. One
corresponds to the so-called hard region, in which all seven alpha parameters
scale as ``1''.  There are twelve so-called soft-collinear regions where
some of the parameters have the weight ``1'' and others the weight ``$\chi$''. For
example, for region 2, we have that
\begin{eqnarray}
\mathrm{region\ 2:}
&\quad&
\left\{
\alpha_1\sim \chi,\,
\alpha_2\sim 1,\,
\alpha_3\sim 1 ,\,
\alpha_4\sim 1 ,\,
\alpha_5\sim \chi ,\,
\alpha_6\sim \chi,\,
\alpha_7\sim 1
\right\}
        \,.
\end{eqnarray}
In total only four regions need to be considered. The remaining
eight regions can be obtained by simple symmetry considerations.

After the expansion, the original integral is expressed as a sum of
homogeneously scaling integrals
\begin{eqnarray}
  J &=&
  {(\mu ^2)^{2\epsilon}}
  e^{2\epsilon \gamma_E}
  \sum_{n=1}^{13}
  \left( \prod_{i=1}^7 \int_0^\infty \alpha_i^{\delta_i} \right)
  \mathcal{U}_{(n)}^{-d/2}
  \exp \left( -\mathcal{F}_{(n)}/\mathcal{U}_{(n)} \right)
  +\mathcal{O}(\chi )
  \,,
\end{eqnarray}
where the summation $n$ spans the relevant regions
and the subscript ``$(n)$'' indicates that the polynomials ${\cal U}$
and ${\cal F}$ from Eq.~(\ref{eq::UF}) are specific to the
corresponding region. They are expanded to leading order in $\chi$.
Note that each integral on the r.h.s. is homogeneous in $m_t^2$ (or
$\chi $) but not in $s$ and $t$, since they are $\mathcal{O}(1)$ parameters.

In the hard region there is only one soft parameter, $m_t^2$, and thus
a naive Taylor expansion of the integrand has to be performed.
This leads to purely massless integrals which are known in the
literature~\cite{Smirnov:1999wz,Bern:2005iz}. We have cross-checked these results
up to the order in $\epsilon$ necessary for our application.
Note that the contribution of the hard region is regular in $\delta_i$
which means that one can take the limit $\delta_i\to 0$ at the very
beginning.

The soft-collinear regions are more involved.  In the following we outline the
calculation of the contribution from region~2 as an example.  The calculation
for other regions proceeds analogously.

\begin{enumerate}
\item We express each integral in terms of two-dimensional
  Mellin-Barnes integrals.  For our example integral, we find the following
  form
\begin{eqnarray}
  J^{(2)} &=&
  {(\mu ^2)^{2\epsilon}}
  e^{2\epsilon \gamma_E}
   \int \frac{dz_1}{2\pi i}\frac{dz_2}{2\pi i}
  \frac{ 
  (-s)^{-\delta_{16}-2}
  (-t)^{-\delta_5-1} (m_t^2)^{-2 \epsilon-\delta_{2347}} 
  \Gamma (\epsilon+\delta_{47}) 
  }
  {\Gamma (\delta_2+1) \Gamma (\delta_3+1) \Gamma
   (\delta_4+1) \Gamma (\delta_7+1) \Gamma (-\epsilon-\delta_{156}-1)}
  \nonumber\\[3mm]
  &&\times
  \frac{
   \Gamma(-z_1)\Gamma (-z_2) 
  \Gamma(z_2-\epsilon-\delta_7+1) 
  \Gamma (z_1-\delta_5+\delta_7)
   \Gamma (z_1-\epsilon -\delta_{156}-1)
   }
   { \Gamma
   (-\epsilon-\delta_{15}+z_1) \Gamma (-\epsilon-\delta_5+z_{12}+1) 
   \Gamma (2\epsilon+\delta_{34}+2\delta_7-z_2)}    
     \nonumber\\[3mm]
  &&\times
  \Gamma (\epsilon+\delta_{37}-z_2)
  \Gamma (\delta_2-\delta_5+z_{12})
   \Gamma (z_{12}-\epsilon-\delta_{15})
   \Gamma (2\epsilon+\delta_{3457}-z_{12})
 \,,\nonumber\\
  \label{eq::2dMB}
\end{eqnarray}
where $z_{12}=z_1+z_2$, $\delta_{123}=\delta_1+\delta_2+\delta_3$ and so on.
Note that the integration contours of $z_1$ and $z_2$
are chosen to be straight lines parallel to the imaginary axis,
satisfying $-1<\mathrm{Re}\,(z_1)<\mathrm{Re}\,(z_2)<0$.

\item Next we use the package {\tt MB.m}~\cite{Czakon:2005rk}
to analytically continue the regularization parameters $\delta_i$
and $\epsilon$ to zero. As a result we obtain two-dimensional
Mellin-Barnes representations which depend only on $z_1, z_2$ and
possibly on $t/s$. Note that the poles in $\delta_i$
cancel among the contributions from the different regions, which provides
a good check of our calculations.

\item We now transform the two-dimensional Mellin-Barnes integrals into
  one-dimensional integrals.  In general this step is non-trivial; we provide more
  details in Appendix~\ref{app::MB2to1}. For simple cases {\tt
    barnesroutines.m}~\cite{barnesroutines} can be used.

\item
At this point we arrive at two types of one-dimensional integral:
those which are independent of $t/s$ and others
which depend on $t/s$, such as
\begin{eqnarray}
\int \frac{dz_1}{2\pi i}
\left(\frac{t}{s}\right)^{z_1}
\Gamma (-1-z_1)^2\Gamma (-z_1)
\Gamma (1+z_1)^2\Gamma (2+z_1)
\psi (1+z_1)\psi '(-z_1)\,.
  \label{eq::1dMB_t}
\end{eqnarray}
We perform a high-precision numerical evaluation (300~digits)
of the $t$-independent integrals and, after summing the contributions from all regions, apply the
PSLQ algorithm~\cite{PSLQ} to re-construct the analytic result as a rational linear
combination of 33 products (up to weight 6) of numbers from the set
\begin{eqnarray}
	\{1,\, \ln 2,\, \pi^2,\, \zeta_3,\, \pi^4,\, \mbox{Li}_4(1/2),\, \zeta_5,\,
	\mbox{Li}_5(1/2),\, \pi^6,\, \mbox{Li}_6(1/2),\, S_{3,3}(-1)\}\,,
\end{eqnarray}
and use a further 200 digits to verify each result. The Nielsen generalized polylogarithm $S_{3,3}(-1)$ is implemented in {\tt Mathematica} as {\tt PolyLog[3,3,-1]}.

For the $t$-dependent integrals
we make an ansatz containing harmonic polylogarithms (HPLs)~\cite{Remiddi:1999ew}
up to weight 6 with alphabet $\nu_i \in \{-1,0\}$,
\begin{align}
\sum _{n=0}^6
c_{\{\nu_i\},n} \mathrm{H}(\{\nu_i\} ,t/s) \log^n(-m_t^2/s)\,,
\end{align}
which we
Taylor-expand in $t/s$.
The series is expressed as a multivariate polynomial
in $(t/s),\log (t/s), \log (-m_t^2/s)$.
We obtain the Taylor series of the integrals by taking their residues at
$z_1=0,1,2,3,...\,$.
We then use 50 low-order terms of the Taylor series to fix the coefficients $c_{\{\nu_i\},n}$ of the ansatz and check
the result using a further 200 higher-order terms of the Taylor series.
\end{enumerate}

Using the above procedure, we obtain the following $\delta_i$-independent result
for our sample master integral $G_{6}(1,1,1,1,1,1,1,0,0)$,
\begin{align}
  J &= -
  {\left(\frac{\mu^2}{-s}\right)^{2\epsilon}
  \frac{1}{s^2 t} }
  \left\{
  l_m^4
  -\frac{8}{3}l_m^3l_t
  +l_m^2\left( 2l_t^2-\frac{2\pi^2}{3}\right)
    +l_m\left(\frac{8\pi^2}{3}l_t-4\zeta_3\right)  
  \right.
  \nonumber\\\vphantom{\bigg\{}
    &
    -\frac{7\pi^4}{15}
    +4l_t \zeta_3
    -2\pi^2l_t^2
    {-}\frac{1}{3}l_t^4
    +8H(-2,0,0;t/s)
    +4\pi^2H(-2;t/s)
  \nonumber\\\vphantom{\bigg\{}
    &
    +\epsilon \left[
    -\frac{4}{3}l_m^5
    +\frac{19}{6}l_m^4l_t
    +l_m^3\left(
    -\frac{4\pi^2}{9}-2l_t^2
    \right)
    \right.
    \nonumber\\\vphantom{\bigg\{}
    &
    +l_m^2\left(
    -4\zeta_3+\pi^2l_t
    +\frac{1}{3}l_t^3
    -\pi^2H(-1,t/s)
    -2H(-1,0,0;t/s)
    \right)
    \nonumber\\\vphantom{\bigg\{}
    &
    +l_m\left(
    -\frac{11\pi^4}{45}
    +8\zeta_3l_t
    -\frac{10\pi^2}{3}l_t^2
    -\frac{2}{3}l_t^4
    +8H(-2,0,0;t/s)+8H(-1,0,0,0;t/s)
    \right.
    \nonumber\\\vphantom{\bigg\{}
    &
    \left.
    +4\pi^2 H(-2;t/s)
    +\frac{4\pi^2}{3} H(-1,0;t/s)
    \right)
    -52\zeta_5+6\pi^2\zeta_3
    +\frac{83\pi^4}{90}l_t
    -4\zeta_3l_t^2
    +\frac{25\pi^2}{9}l_t^3+\frac{l_t^5}{2}
    \nonumber\\\vphantom{\bigg\{}
    &
    -22\pi^2H(-3;t/s)
    -\frac{23\pi^4}{30}H(-1;t/s)
    +16\pi^2H(-2,1;t/s)
    +6\pi^2H(-1,-2;t/s)
    \nonumber\\\vphantom{\bigg\{}
    &
    -44H(-3,0,0;t/s)
    -\frac{10\pi^2}{3}H(-1,0,0;t/s)
    +32H(-2,-1,0,0;t/s)
    \nonumber\\\vphantom{\bigg\{}
    &
    -48H(-2,0,0,0;t/s)+12H(-1,-2,0,0;t/s)
    -12H(-1,0,0,0,0;t/s)
    \nonumber\\\vphantom{\bigg\{}
    &
    \left.
    \left.
    +32\zeta_3H(-2;t/s)
    -24\zeta_3H(-1,0;t/s)
    -\frac{40\pi^2}{3}H(-2,0;t/s)
    \right]
    \right\} + {\cal O}(m_t^2) + {\cal O}(\epsilon^2)\,,
  \label{eq::ana}
\end{align}
where $l_m=\log (-m_t^2/s)$ and $l_t=\log (t/s)$.

To solve the $m_t^2$-differential equation
for most of our master integrals, it is sufficient to obtain just the
leading term in the small-$m_t$ expansion of the boundary condition. However,
for 9 integrals it is also necessary to compute
the next-to-leading term in the asymptotic expansion.
In most of these cases we can simply apply the method discussed above
and expand up to the next-to-leading term in $\chi$. However,
for two of the seven-line integrals,
Step~3 above is hard to apply at the next-to-leading order.
For these integrals, we use the corresponding $t$-differential equations
to obtain the next-to-leading boundary conditions.

For example, the next-to-leading term of $G_{6}(1,1,1,1,1,1,2,0,0)$
is determined in the following way:
we consider the $t$-differential equation
of $G_{6}(1,1,1,1,1,1,1,0,0)$, which is fully known at the leading order,
\begin{eqnarray}
  \label{eq::g6diffeq}
  \frac{\mbox{d}}{\mbox{d}t}G_{6}(1,1,1,1,1,1,1,0,0)
    &=&+ A \: G_{6}(1,1,1,1,1,1,1,0,0)\nonumber\\
    && + B \: m_t^2 \: G_{6}(1,1,1,1,1,2,1,0,0)\nonumber\\
    && + C \: G_{6}(1,1,1,1,1,1,2,0,0)\nonumber\\
    && + (\mbox{known lower-line integrals})\,
\end{eqnarray}
where $A$, $B$ and $C$ are $\mathcal{O}(1)$ coefficients.
The leading terms of the integrals behave as follows,
\begin{eqnarray}
G_{6}(1,1,1,1,1,1,1,0,0) &=& \mathcal{O}(1)\nonumber\\
G_{6}(1,1,1,1,1,2,1,0,0) &=& \mathcal{O}(1/m_t^2)\nonumber\\
G_{6}(1,1,1,1,1,1,2,0,0) &=& \mathcal{O}(1/m_t^2)\,.
\end{eqnarray}
Therefore, the next-to-leading ($\mathcal{O}(1)$) contribution to
$G_{6}(1,1,1,1,1,1,2,0,0)$ appears in the differential equation at
$\mathcal{O}(1)$ alongside the (known) leading contributions to
$G_{6}(1,1,1,1,1,1,1,0,0)$ and $G_{6}(1,1,1,1,1,2,1,0,0)$ and can be easily
determined.
Note that the leading order ($\mathcal{O}(1/m_t^2)$) contribution of
$G_{6}(1,1,1,1,1,1,2,0,0)$ cancels against $\mathcal{O}(1/m_t^2)$ terms
from the lower-line integrals appearing in Eq.~(\ref{eq::g6diffeq}).

We have checked all boundary conditions numerically at a few euclidian
values of $s$ and $t$ using the command {\tt SDExpandAsy} of {\tt
  FIESTA}~\cite{Smirnov:2015mct}.

\subsection{Solving the differential equations}

Using the boundary conditions discussed in the previous subsection we can
solve the differential equations of Subsection~\ref{sub::diffeq} in a
straightforward way. All of our results are expressed in terms of HPLs.  The
expansion depth is limited only by the size of the intermediate
expressions which enter the system of linear equations for the coefficients in
our ansatz. We have expanded each master integral such that the final result
for the amplitude $gg\to HH$ is available up to order $m_t^{16}$.
Our results for the master integrals can be downloaded from~\cite{progdata}.

\begin{figure}[t]
  \centering
  \begin{tabular}{cc}
    \mbox{}\hspace*{-2.5em}\mbox{}
    \includegraphics[width=0.55\textwidth]{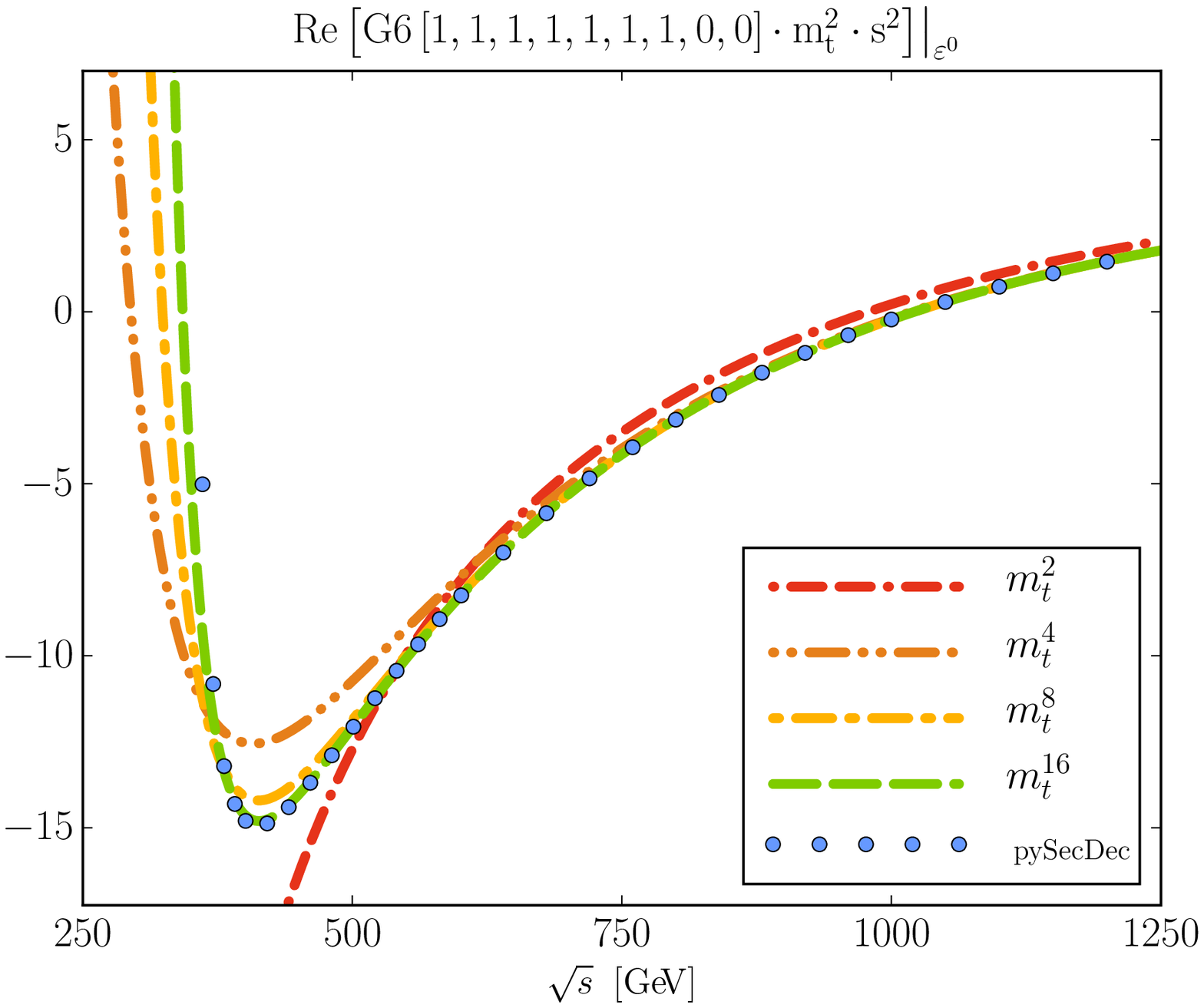} &
    \mbox{}\hspace*{-2.5em}\mbox{}
    \includegraphics[width=0.55\textwidth]{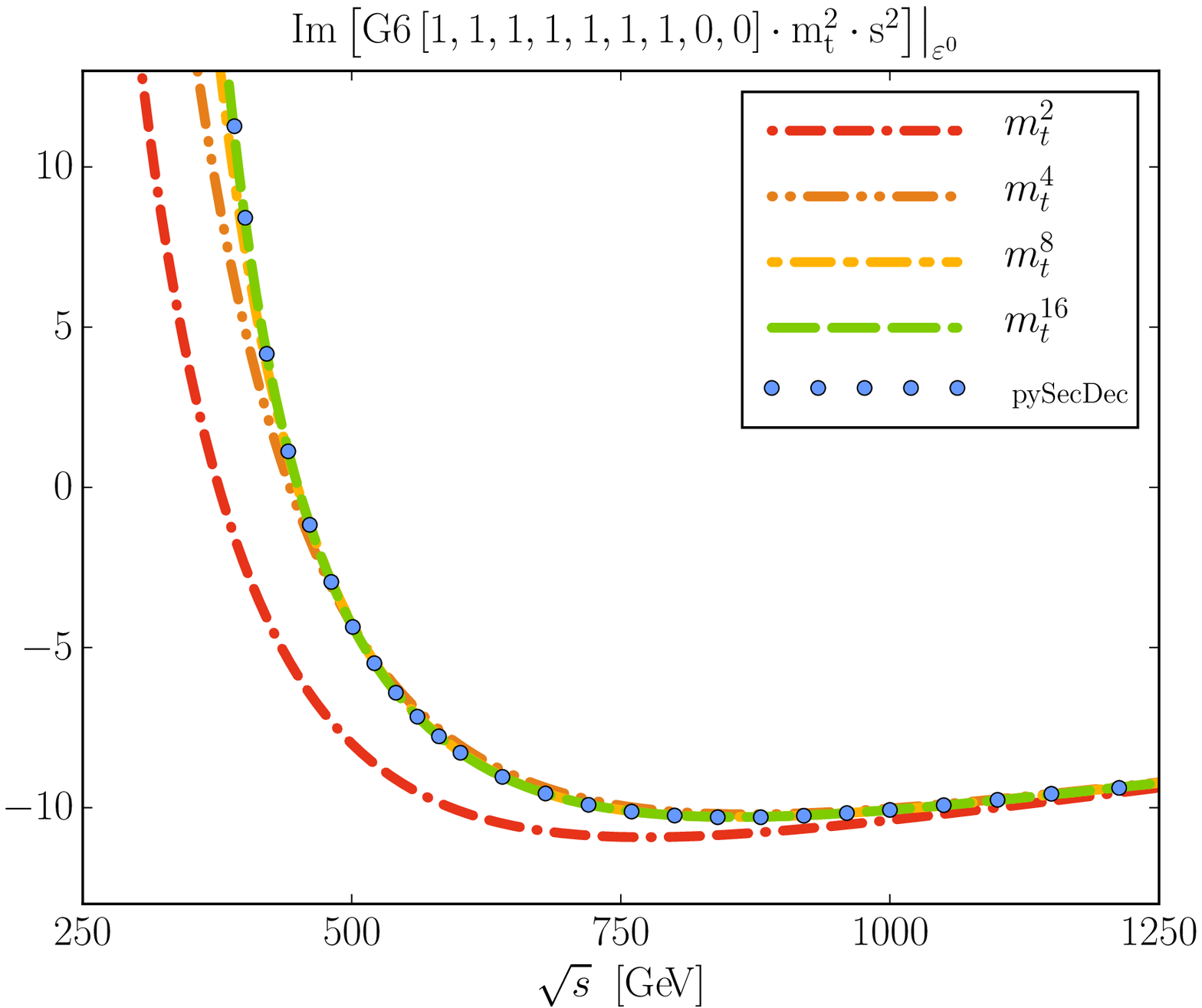}
    \\
    \mbox{}\hspace*{-2.5em}\mbox{}
    \includegraphics[width=0.55\textwidth]{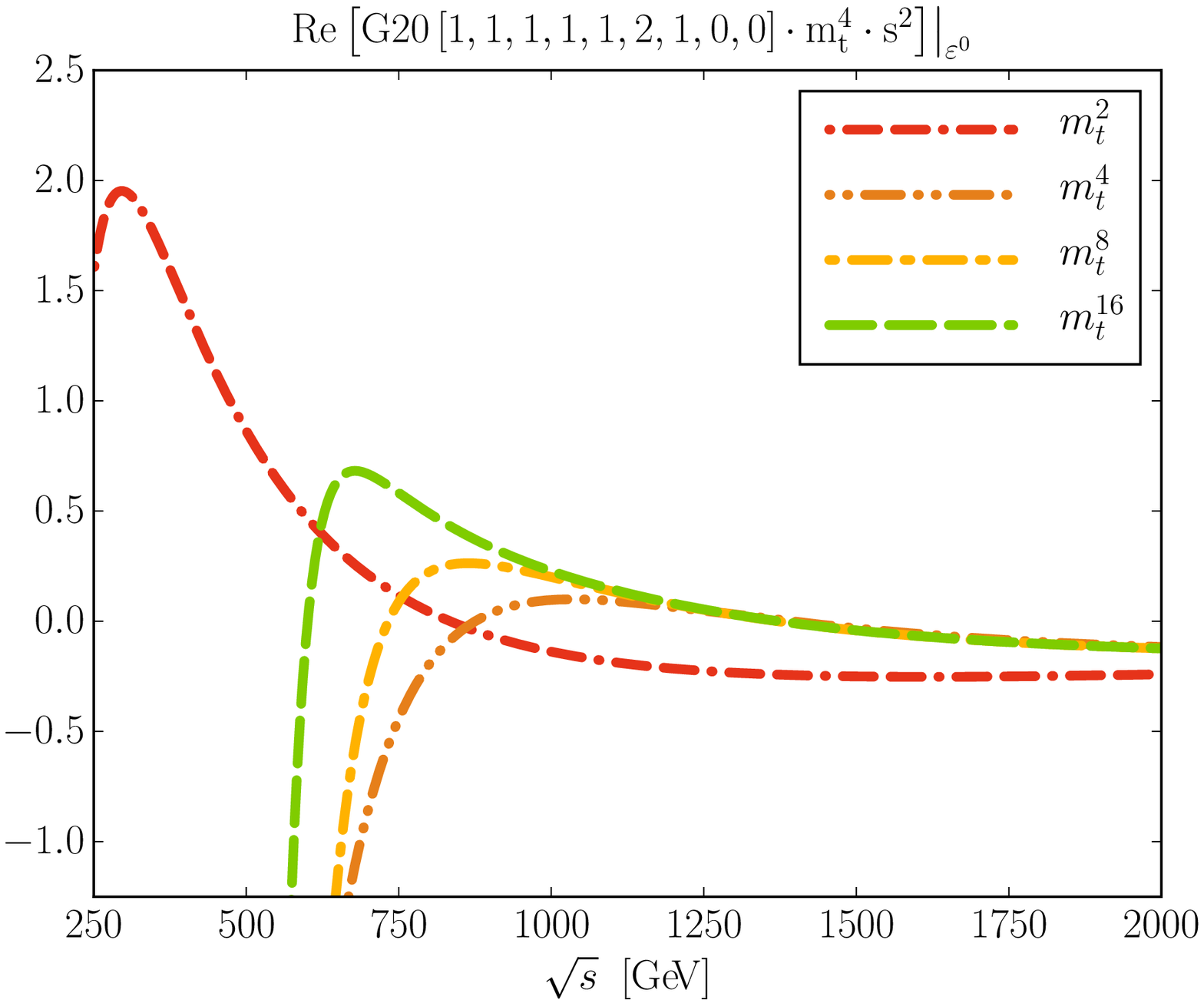} &
    \mbox{}\hspace*{-2.5em}\mbox{}
    \includegraphics[width=0.55\textwidth]{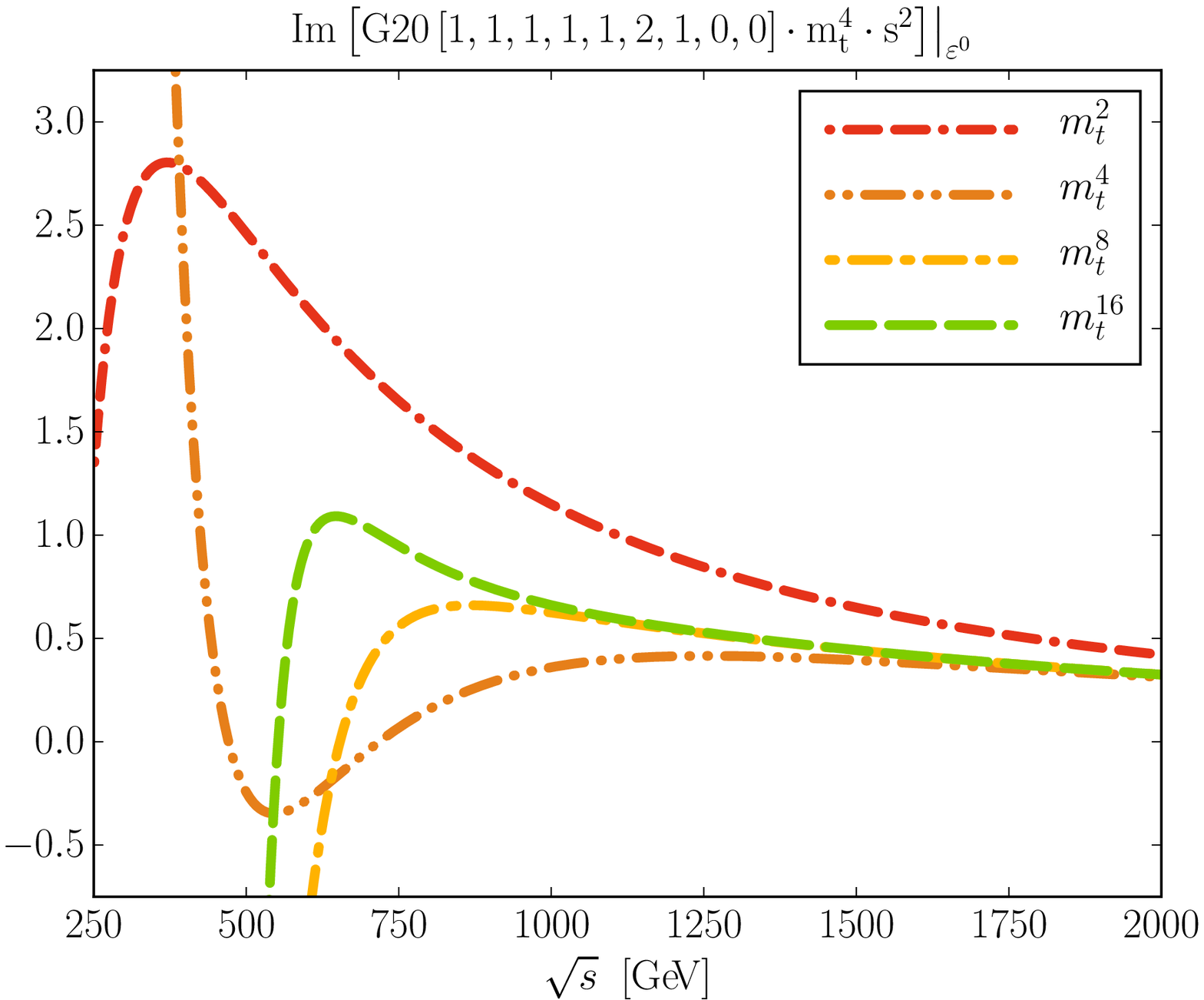}
  \end{tabular}
  \caption{\label{fig::MIs_s}Real and imaginary part of the $\epsilon^0$ term
    of the two master integrals $G_{6}(1,1,1,1,1,1,1,0,0)$ and
    $G_{20}(1,1,1,1,1,2,1,0,0)$.  For convenience we multiply by powers of
    $m_t$ and $s$ as indicated above the plot.}
\end{figure}

For illustration we show in Fig.~\ref{fig::MIs_s} the results for two master
integrals: $G_{6}(1,1,1,1,1,1,1,0,0)$, which is used as an example in
Subsection~\ref{sub::BC}, and $G_{20}(1,1,1,1,1,2,1,0,0)$. We plot the real
and imaginary parts of the $\epsilon^0$ term as a function of $\sqrt{s}$ and
choose $t=-s/2$, which corresponds to $\theta=\pi/2$
(cf. Fig.~\ref{fig::ds_s}), $m_t=175$~GeV and $\mu^2=s$.  For clarity we
multiply each integral by appropriate powers of $m_t$ and $s$ such that the
leading term starts with $m_t^2$ and is dimensionless.  In each case we
display the approximations including $m_t^2$, $m_t^4$, $m_t^8$ and $m_t^{16}$
terms.

In the panels showing $G_{6}(1,1,1,1,1,1,1,0,0)$ we compare the approximations
to the exact result, which has been obtained numerically using {\tt
  pySecDec}~\cite{Borowka:2017idc}.  For this integral we observe a rapid
convergence. In fact, the $m_t^4$, $m_t^8$ and $m_t^{16}$ curves agree with
each other and the exact points down to $\sqrt{s} \approx 600$~GeV and the two
highest approximation even down to $\sqrt{s} \approx 500$~GeV. It is
interesting to note that the $m_t^{16}$ curve reproduces to high 
accuracy the turning point at $\sqrt{s} \approx 400$~GeV and the steep rise
below that energy. It can not be expected that all master integrals show such
good convergence properties. In fact there are integrals, in particular some
of the non-planar contributions, where the expansion parameter is $m_t^2/u$
instead of $m_t^2/s$ which results in a smaller radius of convergence.

For $G_{20}(1,1,1,1,1,2,1,0,0)$ we were not able to obtain stable numerical
results using {\tt pySecDec} and thus we only show our
approximations. We observe a similar pattern as for the LO cross section shown in
Fig.~\ref{fig::ds_s}: the inclusion of more terms extends the convergence range in
$\sqrt{s}$ down to smaller values. Furthermore, the curves including $m_t^8$ and $m_t^{16}$
terms agree down to $\sqrt{s}\approx 900$~GeV and it can be expected that above
this energy a good approximation to the exact result can be provided.


\section{Conclusions}

The main focus of this paper is on NLO corrections to double Higgs boson
production in the high energy region, where the top quark mass is assumed to be
small compared to the kinematic variables $s$ and $t$. Such considerations
complement expansions for large top quark mass and around the threshold which have
been considered in the literature, see
Refs.~\cite{Grigo:2013rya,Grober:2017uho}.  Furthermore, they provide an
indpendent cross check of the exact calculation~\cite{Borowka:2016ypz} which
relies heavily on numerical methods.

In this paper we perform the reduction of the $gg\to HH$ amplitude to master
integrals and compute all planar integrals in an expansion for small
$m_t^2$. The expansion depth for each master integral is chosen such that the
amplitude includes terms up to order $m_t^{16}$. Our analytic results for
the master integrals are expressed in terms of HPLs and can be obtained in
computer-readable form from~\cite{progdata}.


 
\section*{Acknowledgements}

This work is supported by the BMBF grant 05H15VKCCA.  D.W. acknowledges the
support by the DFG-funded Doctoral School KSETA.
We thank Alexander Smirnov for providing us with unpublished versions of {\tt
  FIRE} which we could use to help optimize our reduction. We thank Alexander
Smirnov and Vladimir Smirnov for many useful discussions. 



\begin{appendix}


\section{\label{app::MIs}One- and two-loop master integrals}

We define the propagators of the one-loop integral family as
\begin{align}
D_1(q_1,q_2,q_3,q_4)=&
\left\{
m_t^{2}-l_1^2,
m_t^{2}-(l_1+q_3)^2,
m_t^{2}-(l_1-q_1-q_2)^2,
m_t^{2}-(l_1-q_1)^2
\right\}
\,,
\end{align}
and at two loops we introduce
\begin{align}
D_{6}(q_1,q_2,q_3,q_4)&=\left\{
m_t^{2}-l_1^2,
m_t^{2}-l_2^2,
m_t^{2}-(l_2+q_3)^2,
m_t^{2}-(l_2-q_1-q_2)^2,
\right.\nonumber\\&\left.
m_t^{2}-(l_1-q_1-q_2)^2,
m_t^{2}-(l_1-q_1)^2,
-(l_1-l_2)^2,
-(l_1+q_3)^2,
\right.\nonumber\\&\left.
-(l_2+q_1)^2
\right\}\,,\nonumber\\
D_{20}(q_1,q_2,q_3,q_4)&=\left\{
-l_1^2,
m_t^{2}-l_2^2,
m_t^{2}-(l_2+q_3)^2,
m_t^{2}-(l_2-q_1-q_2)^2,
\right.\nonumber\\
&\left.
-(l_1-q_1-q_2)^2,
-(l_1-q_1)^2,
m_t^{2}-(l_1-l_2)^2,
-(l_1+q_3)^2,
-(l_2+q_1)^2
\right\}\,,\nonumber\\
D_{33}(q_1,q_2,q_3,q_4)&=\left\{
-l_1^2,
m_t^{2}-l_2^2,
m_t^{2}-(l_2+q_4)^2,
-(l_1+q_3+q_4)^2,
-(l_1-q_1)^2,
\right.\nonumber\\
&\left.
m_t^{2}-(l_1-l_2+q_3)^2,
m_t^{2}-(l_1-l_2)^2,
-(l_1+q_4)^2,
-(l_2+q_1)^2
\right\}\,,\nonumber\\
D_{47}(q_1,q_2,q_3,q_4)&=\left\{
-l_1^2,
m_t^{2}-l_2^2,
m_t^{2}-(l_2+q_4)^2,
m_t^{2}-(l_2-q_1-q_2)^2,
\right.\nonumber\\&\left.
m_t^{2}-(l_1-l_2+q_2)^2,
m_t^{2}-(l_1-l_2)^2,
-(l_1-q_1)^2,
-(l_1+q_4)^2,
\right.\nonumber\\&\left.
-(l_2+q_1)^2
\right\}\,,\nonumber\\
D_{72}(q_1,q_2,q_3,q_4)&=\left\{
m_t^{2}-l_1^2,
m_t^{2}-(l_1+q_2)^2,
m_t^{2}-(l_1+q_1+q_2)^2,
m_t^{2}-(l_2+q_1+q_2)^2,
\right.\nonumber\\
&\left.
m_t^{2}-(l_2-q_3)^2,
m_t^{2}-(l_1-q_3)^2,
-(l_1-l_2)^2,
-(l_2+q_2)^2,
-(l_2+q_3)^2
\right\}\,,\nonumber\\
D_{75}(q_1,q_2,q_3,q_4)&=\left\{
m_t^{2}-l_1^2,
m_t^{2}-(l_1+q_4)^2,
m_t^{2}-(l_1-q_1-q_2)^2,
-(l_2-q_1-q_2)^2,
\right.\nonumber\\
&\left.
-(l_2-q_1)^2,
m_t^{2}-(l_1-q_1)^2,
m_t^{2}-(l_1-l_2)^2,
-(l_2+q_4)^2,
-(l_2+q_1)^2
\right\}\,,\nonumber\\
D_{90}(q_1,q_2,q_3,q_4)&=\left\{
m_t^{2}-l_1^2,
m_t^{2}-(l_1+q_3)^2,
-(l_1+l_2-q_1-q_2)^2,
-(l_1+l_2-q_1)^2,
\right.\nonumber\\
&\left.
m_t^{2}-(l_1-q_1)^2,
m_t^{2}-(l_2+q_4)^2,
m_t^{2}-l_2^2,
-(l_2+q_3)^2,
-(l_2+q_1)^2
\right\}\,,\nonumber
\end{align}
where $l_1$ and $l_2$ are the loop momenta.  Note that we prefer to use
our internal notation for the families, which is the reason why the
numbering is not sequential.  We also define the integral families which
are obtained by the exchange of external momenta.  At one-loop order
there are two more families which are related to $D_{1}(q_1,q_2,q_4,q_3)$ as
follows,
\begin{align}
\begin{array}{ll}
D_{2}(q_1,q_2,q_3,q_4)=D_{1}(q_1,q_2,q_4,q_3)&
D_{3}(q_1,q_2,q_3,q_4)=D_{1}(q_1,q_4,q_3,q_2)\,.\\
\end{array}
\end{align}
At two-loop order we have
\begin{align}
\begin{array}{ll}
D_{4}(q_1,q_2,q_3,q_4)=D_{6}(q_1,q_4,q_3,q_2)&
D_{5}(q_1,q_2,q_3,q_4)=D_{6}(q_1,q_2,q_4,q_3)\\
D_{8}(q_1,q_2,q_3,q_4)=D_{6}(q_4,q_1,q_3,q_2)&
D_{10}(q_1,q_2,q_3,q_4)=D_{6}(q_3,q_1,q_4,q_2)\\
D_{11}(q_1,q_2,q_3,q_4)=D_{6}(q_3,q_1,q_2,q_4)&
D_{26}(q_1,q_2,q_3,q_4)=D_{20}(q_4,q_3,q_1,q_2)\\
D_{51}(q_1,q_2,q_3,q_4)=D_{47}(q_2,q_1,q_3,q_4)&
D_{59}(q_1,q_2,q_3,q_4)=D_{47}(q_2,q_3,q_1,q_4)\\
D_{71}(q_1,q_2,q_3,q_4)=D_{72}(q_1,q_2,q_4,q_3)&
D_{73}(q_1,q_2,q_3,q_4)=D_{72}(q_1,q_4,q_2,q_3)\\
D_{78}(q_1,q_2,q_3,q_4)=D_{75}(q_4,q_1,q_2,q_3)&
D_{79}(q_1,q_2,q_3,q_4)=D_{75}(q_2,q_1,q_3,q_4)\\
D_{84}(q_1,q_2,q_3,q_4)=D_{75}(q_3,q_2,q_4,q_1)&
D_{91}(q_1,q_2,q_3,q_4)=D_{90}(q_4,q_1,q_2,q_3)\,.
\end{array}
\end{align}

Our minimal set of one-loop master integrals is given by
\begin{align*}
\begin{array}{lllll}
G_1(1,1,1,1),&G_2(1,0,1,0),&G_2(1,1,1,0),&G_2(1,1,1,1),&G_3(0,0,0,1),\\
G_3(0,1,0,1),&G_3(1,0,1,0),&G_3(1,1,0,1),&G_3(1,1,1,0),&G_3(1,1,1,1),
\end{array}
\end{align*}
and at two loops we have
{\scalefont{0.75}
\begin{align}
\begin{array}{lllll}
G_{4}(1,1,1,1,1,1,1),&G_{4}(1,1,1,1,1,1,2),&G_{4}(1,1,1,1,1,2,1),&G_{5}(1,1,1,1,1,1,1),&G_{5}(1,1,1,1,1,1,2),\\
G_{5}(1,1,1,1,1,2,1),&G_{6}(1,1,0,1,1,0,0),&G_{6}(1,1,0,1,1,1,0),&G_{6}(1,1,1,1,1,1,0),&G_{6}(1,1,1,1,1,1,1),\\
G_{6}(1,1,1,1,1,1,2),&G_{6}(1,1,1,1,1,2,1),&G_{8}(1,1,0,1,1,0,0),&G_{8}(1,1,0,1,1,1,0),&G_{8}(1,1,1,1,1,1,0),\\
G_{8}(1,1,1,1,1,1,1),&G_{8}(1,1,1,1,1,1,2),&G_{8}(1,1,1,1,1,2,1),&G_{10}(1,1,1,1,1,1,1),&G_{10}(1,1,1,1,1,1,2),\\
G_{10}(1,1,1,1,1,2,1),&G_{11}(1,1,0,1,1,0,0),&G_{11}(1,1,0,1,1,1,0),&G_{11}(1,1,1,1,1,1,0),&G_{11}(1,1,1,1,1,1,1),\\
G_{11}(1,1,1,1,1,1,2),&G_{11}(1,1,1,1,1,2,1),&G_{20}(1,1,0,1,1,0,0),&G_{20}(1,1,0,1,1,0,1),&G_{20}(1,1,1,1,1,0,0),\\
G_{20}(1,1,1,1,1,0,1),&G_{20}(1,1,1,1,1,1,1),&G_{20}(1,1,1,1,1,1,2),&G_{20}(1,1,1,1,1,2,1),&G_{20}(1,1,2,1,1,1,1),\\
G_{26}(1,1,1,1,1,1,1),&G_{26}(1,1,1,1,1,1,2),&G_{26}(1,1,1,1,1,2,1),&G_{26}(1,1,2,1,1,1,1),&G_{33}(1,0,1,1,1,0,1),\\
G_{33}(1,0,1,1,2,0,1),&G_{33}(1,1,0,1,0,0,0),&G_{33}(1,1,0,1,0,1,0),&G_{33}(1,1,0,1,0,2,0),&G_{33}(1,1,0,1,1,1,0),\\
G_{33}(1,1,0,1,2,1,0),&G_{33}(1,1,1,1,0,1,0),&G_{33}(1,1,1,1,0,1,1),&G_{33}(1,1,1,1,0,2,1),&G_{33}(1,1,1,1,1,0,1),\\
G_{33}(1,1,1,1,1,1,0),&G_{33}(1,1,1,1,1,1,1),&G_{33}(1,1,1,1,2,1,1),&G_{33}(1,1,1,2,1,1,1),&G_{33}(1,1,2,1,1,1,1),\\
G_{47}(1,0,1,1,1,2,1),&G_{47}(1,1,1,0,1,2,1),&G_{47}(1,1,1,1,1,1,2),&G_{47}(1,1,1,2,1,0,0),&G_{47}(1,1,1,2,1,1,1),\\
G_{51}(1,1,0,1,1,1,1),&G_{51}(1,1,1,1,1,1,1),&G_{51}(1,1,1,1,1,2,1),&G_{51}(1,1,1,1,2,1,1),&G_{59}(1,0,1,1,1,1,1),\\
G_{59}(1,1,0,1,1,1,1),&G_{59}(1,1,1,1,1,1,1),&G_{59}(1,1,1,1,1,2,1),&G_{59}(1,1,1,1,2,1,1),&G_{59}(1,2,1,1,1,1,1),\\
G_{59}(2,1,1,1,1,1,1),&G_{71}(1,1,1,1,1,0,1),&G_{71}(1,1,2,1,1,0,1),&G_{71}(1,2,0,1,0,1,1),&G_{72}(0,1,0,1,1,2,1),\\
G_{72}(1,0,1,1,1,0,1),&G_{72}(1,0,2,1,1,0,1),&G_{72}(1,1,0,1,1,0,1),&G_{72}(1,1,0,1,1,0,2),&G_{72}(1,1,0,1,1,0,3),\\
G_{72}(1,1,0,1,1,1,1),&G_{72}(1,1,0,1,1,2,1),&G_{72}(1,1,1,1,1,0,1),&G_{72}(1,1,2,1,1,0,1),&G_{73}(0,1,0,1,0,1,2),\\
G_{73}(0,1,0,1,1,2,1),&G_{73}(0,1,1,1,1,0,1),&G_{73}(1,0,1,1,1,0,1),&G_{73}(1,1,0,1,1,2,1),&G_{73}(1,1,1,1,1,0,1),\\
G_{73}(1,1,2,1,1,0,1),&G_{75}(1,1,0,1,0,2,1),&G_{78}(1,1,0,1,0,1,1),&G_{78}(1,1,0,1,0,2,1),&G_{78}(1,1,1,0,0,1,1),\\
G_{78}(1,1,1,0,1,0,1),&G_{78}(1,2,1,0,1,0,1),&G_{78}(2,1,0,1,0,1,1),&G_{79}(1,0,1,0,0,0,1),&G_{79}(1,0,1,0,0,1,1),\\
G_{79}(1,0,1,0,1,0,1),&G_{79}(1,0,1,0,1,0,2),&G_{79}(1,0,1,0,2,0,1),&G_{79}(1,1,1,0,0,1,1),&G_{79}(1,1,1,0,1,0,1),\\
G_{79}(1,2,1,0,1,0,1),&G_{84}(0,1,0,0,2,0,1),&G_{84}(1,0,1,0,0,0,1),&G_{84}(1,0,1,0,0,1,1),&G_{84}(1,0,1,0,1,0,1),\\
G_{84}(1,0,1,0,1,0,2),&G_{84}(1,0,1,0,2,0,1),&G_{84}(1,1,1,0,0,1,1),&G_{84}(1,1,1,0,1,0,1),&G_{84}(1,2,1,0,1,0,1),\\
G_{84}(2,1,1,0,1,0,1),&G_{90}(1,1,1,0,1,1,1),&G_{90}(1,1,1,1,2,1,1),&G_{90}(1,1,1,2,1,1,1),&G_{90}(1,2,1,0,1,1,1),\\
G_{91}(0,0,0,0,1,0,1),&G_{91}(0,0,1,0,1,1,0),&G_{91}(0,0,1,0,1,1,1),&G_{91}(0,1,0,0,1,0,1),&G_{91}(0,1,0,1,1,1,0),\\
G_{91}(0,1,0,1,1,1,1),&G_{91}(0,1,0,1,1,2,0),&G_{91}(0,1,1,1,1,1,1),&G_{91}(1,0,0,1,0,1,0),&G_{91}(1,0,0,1,0,1,1),\\
G_{91}(1,0,0,1,1,1,1),&G_{91}(1,0,0,2,0,1,0),&G_{91}(1,0,1,0,0,0,1),&G_{91}(1,0,1,0,0,1,1),&G_{91}(1,0,1,0,1,1,1),\\
G_{91}(1,0,1,1,0,1,1),&G_{91}(1,0,1,1,1,1,0),&G_{91}(1,0,1,1,1,1,1),&G_{91}(1,0,1,1,1,1,2),&G_{91}(1,0,1,1,1,2,0),\\
G_{91}(1,0,2,0,0,0,1),&G_{91}(1,0,2,0,1,1,1),&G_{91}(1,0,3,0,1,1,1),&G_{91}(1,1,0,0,1,0,1),&G_{91}(1,1,0,1,0,1,1),\\
G_{91}(1,1,0,1,1,1,0),&G_{91}(1,1,0,2,0,1,1),&G_{91}(1,1,0,3,0,1,1),&G_{91}(1,1,1,0,0,1,1),&G_{91}(1,1,1,0,1,0,1),\\
G_{91}(1,1,1,0,1,1,1),&G_{91}(1,1,1,1,0,0,1),&G_{91}(1,1,1,1,0,0,2),&G_{91}(1,1,1,1,0,1,1),&G_{91}(1,1,1,1,1,1,1),\\
G_{91}(1,1,1,1,1,1,2),&G_{91}(1,1,1,1,1,2,1),&G_{91}(1,2,1,0,1,0,1),&G_{91}(2,0,1,1,0,1,1),&G_{91}(2,1,0,1,1,1,0),\\
G_{91}(2,1,1,0,1,0,1).&&&&
\end{array}
\nonumber
\end{align}
}
\vspace{-5mm}
\begin{equation}
\label{eq::MI2l}
{}
\end{equation}

\begin{figure}[t]
  \centering
  \includegraphics[width=\textwidth]{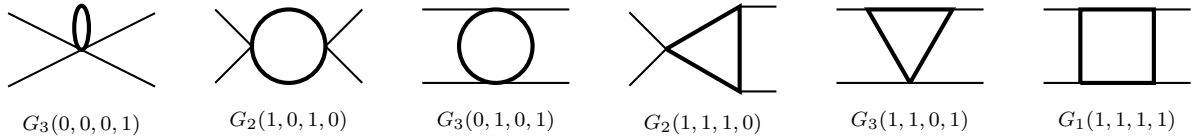}
  \caption{\label{fig::mi1}One-loop master integrals. Solid and dashed lines
    represent massive and massless scalar propagators, respectively. The external (thin) lines are
    massless. The four master integrals which are not shown are
      obtained by crossing.}
\end{figure}

\begin{figure}[t]
  \centering
  \includegraphics[width=\textwidth]{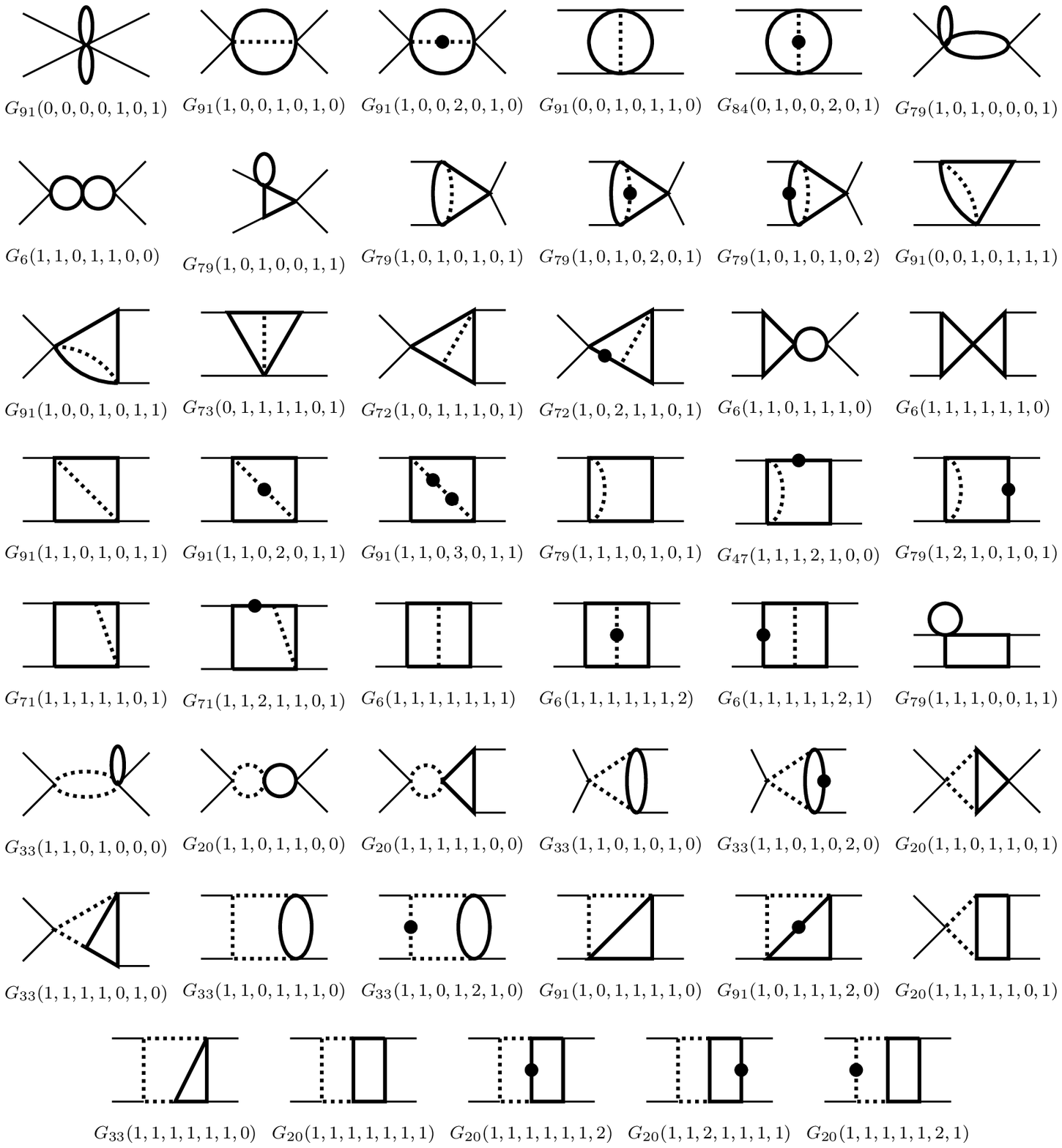}
  \caption{\label{fig::mi2}Two-loop planar 
    master integrals. Solid and dashed lines
    represent massive and massless scalar propagators, respectively. The
    external (thin) lines are massless. The planar master integrals 
    form~(\ref{eq::MI2l}) which are not shown are obtained by crossing.}
\end{figure}

Note that at two-loop order, each family is defined using nine
propagators.  However, in our case the master integrals are chosen such, that
the last two indices are always zero. Thus, we have omitted them in the above
list.  For convenience we show in Figs.~\ref{fig::mi1} and~\ref{fig::mi2}
graphical representations of our one- and two-loop master integrals.

We have cross-checked our expressions against the results given in the
ancillary file of~\cite{Kudashkin:2017skd} and find complete agreement.  Note
that in~\cite{Kudashkin:2017skd} the master integrals are expanded to a
sufficient order to provide an amplitude to ${\cal O}(m_t^{2})$. Here, we
compute each master integral to a sufficient depth to provide the $gg\to HH$
amplitude at ${\cal O}(m_t^{16})$.  We have successfully compared the
triangle master integrals to Ref.~\cite{Anastasiou:2006hc}.  In the
ancillary file to this paper~\cite{progdata} we provide analytic results for
all one- and two-loop planar master integrals discussed here.  Note that as
the integration measure we use $(\mu^2)^{(4-d)/2}e^{\epsilon\gamma_E}{\rm
  d}^d k/{ (i\pi^{d/2}) }$ where $d=4-2\epsilon$ is the space-time
dimension.


\section{\label{app::MB2to1}Reducing the dimensionality of Mellin-Barnes integrals}

In this appendix, we consider the following two types of Mellin-Barnes integrals:
\begin{align}
	&
	\int_{C}^{}
	\frac{dz}{2\pi i}
	\Gamma
	\left[
	\begin{array}[]{c}
		a_1-z,a_2-z,b_1+z,b_2+z,b_3+z\\
		c+z
	\end{array}
	\right],
	\label{appB:eq1}
	\\
	&
	\int_{C}^{}
	\frac{dz}{2\pi i}
	\Gamma
	\left[
	\begin{array}[]{c}
		a_1-z,a_2-z,b_1+z,b_2+z,b_3+z\\
		c+z
	\end{array}
	\right]
	\psi (X)
	\label{appB:eq2}
	\,,
\end{align}
with $X=a_1-z,a_2-z,b_1+z,b_2+z,b_3+z$ or $c+z$, where
the following compact notation has been introduced
\begin{align}
	\Gamma\left[ x_1,\dots,x_n \right]
	=\prod_{i=1}^n \Gamma (x_i)
	\, ,\qquad
	\Gamma\left[ 
		\begin{array}[]{c}
			x_1,\dots,x_n\\
			y_1,\dots,y_m
		\end{array}
	\right]
	=\frac{\Gamma\left[ x_1,\dots,x_n \right]}
	{\Gamma\left[ y_1,\dots,y_m \right]}
	\,.
	\label{appB:notations}
\end{align}
Furthermore, we use the compressed notation $a_{12}=a_1+a_2,
b_{123}=b_1+b_2+b_3$.  In Eqs.~(\ref{appB:eq1}) and~(\ref{appB:eq2}) the
integration contour $C$ goes from $-i\infty$ to $+i\infty$ and it is assumed
that all the poles of $\Gamma\left[a_1-z,a_2-z\right]$ lie to the right of $C$
and those of $\Gamma\left[b_1+z,b_2+z,b_3+z\right]$ to the left of $C$ in the
complex $z$-plane.  We also assume
\begin{align}
  \mathrm{Re}\, (a_i)>0, 
  \: \mathrm{Re}\, (b_j)>0
  \:\: \mbox{for all} 
  \:\: i,j,
  \label{appB::assumptions}
\end{align}
and choose the contour to be along the imaginary axis.  If some of the left
poles and the right poles merge, a regularization and a subsequent analytic
continuation are required (see the example, discussed below
Eq.~\eqref{appB:example}).

Let us first briefly summarize the known properties of Mellin-Barnes integrals.
\begin{itemize}
\item
  If $c=b_3$,
  the first Barnes lemma 
  \begin{align}
    &\int_{C}^{}
    \frac{dz}{2\pi i}
    \Gamma\left[ 
      a_1-z,a_2-z,b_1+z,b_2+z
    \right]
    =\Gamma\left[ 
      \begin{array}[]{c}
        a_1+b_1,a_1+b_2,a_2+b_1,a_2+b_2\\
        a_1+a_2+b_1+b_2
      \end{array}
    \right]
    ,
    \label{appB:i1}	
  \end{align}
  can be applied to Eq.~\eqref{appB:eq1}. By taking a derivative
  w.r.t. one of the parameters (e.g.~$a_1$) one obtains a solution for
  Eq.~\eqref{appB:eq2}.
      
\item
  If $c=a_{12}+b_{123}$,
  the second Barnes lemma 
  \begin{align}
    &
    \int_{C}^{}
    \frac{dz}{2\pi i}
    \Gamma\left[ 
      \begin{array}[]{c}
        a_1-z,a_2-z,b_1+z,b_2+z,b_3+z\\
        a_1+a_2+b_1+b_2+b_3+z
      \end{array}
    \right]
    \nonumber\\& \hspace{2cm}
    =\Gamma\left[ 
      \begin{array}[]{c}
        a_1+b_1,a_1+b_2,a_1+b_3,a_2+b_1,a_2+b_2,a_2+b_3\\
        a_{12}+b_{12},a_{12}+b_{13},a_{12}+b_{23}
      \end{array}
    \right]
    ,
    \label{appB:l2}
  \end{align}
  can be applied to Eq.~\eqref{appB:eq1}.
  However, there is no corresponding expression for Eq.~\eqref{appB:eq2};
  differentiation w.r.t. the parameters $a_i,b_j$
  gives relations between expressions which have the form of Eq.~\eqref{appB:eq2}
  with different $X=a_1-z,a_2-z,b_1+z,b_2+z,b_3+z,c+z$.
  The number of independent relations is
  smaller than the number of possible choices of $X$. Thus, no analytic result
  for Eq.~\eqref{appB:eq2} can be obtained.

\item 
  If $c=a_{12}+b_{123}+1$ or $c=a_{12}+b_{123}+2$
  the solutions for Eq.~\eqref{appB:eq1} are also known \cite{Smirnov:2012gma}
  and implemented in the package \texttt{barnesroutines.m}~\cite{barnesroutines}.
\end{itemize}

In the following we sketch the derivation of a solution for Eq.~\eqref{appB:eq1}
for the general case, which also yields a solution for
Eq.~\eqref{appB:eq2} after differentiation w.r.t. one of the parameters.

Based on the assumptions about the relation between $a_i,b_j$ (Eq.~(\ref{appB::assumptions})),
we can express Eq.~\eqref{appB:eq1} as
\begin{align}
	\int_{C}^{}
	\frac{dz}{2\pi i}
	\Gamma
	\left[
	\begin{array}[]{c}
		a_1-z,a_2-z,b_1+z,b_2+z,b_3+z\\
		c+z
	\end{array}
	\right]
	=-\sum_{m=0}^{\infty}
	\left( \mathrm{Res}_{z=a_1+m}
	+\mathrm{Res}_{z=a_2+m}\right),
	\label{appB:der1}
\end{align}
where
\begin{align}
	-&\sum_{m=0}^{\infty}
	\mathrm{Res}_{z=a_1+m}
	\nonumber\\
	&=
	\sum_{m=0}^{\infty}
	(-1)^m 
	\Gamma
	\left[
	\begin{array}[]{c}
		-a_1+a_2-m,a_1+b_1+m,a_1+b_2+m,a_1+b_3+m\\
		m+1,a_1+c+m
	\end{array}
	\right]
	\nonumber\\
	&=
	\Gamma
	\left[
	\begin{array}[]{c}
		-a_1+a_2,a_1+b_1,a_1+b_2,a_1+b_3\\
		a_1+c
	\end{array}
	\right]
	\, _3F_2
	\left(  
	\begin{array}{c}
		a_1+b_1,a_1+b_2,a_1+b_3\\
		1+a_1-a_2,a_1+c
	\end{array};1
	\right)\,.
	\label{}
\end{align}
$_3F_2$ is the generalized hypergeometric function.
The residues at $z=a_2+m$
are written in a similar manner.
At this point the r.h.s. of Eq.~\eqref{appB:der1}
contains two $_3F_2$.
We can transform it into an expression
containing only one $_3F_2$
using the relation~\cite{table},
\begin{align}
	_3F_2
	\left( 
	\begin{array}[]{c}
		a,b,c\\
		d,e
	\end{array};1
	\right)
	=&\:
	\Gamma
	\left[
	\begin{array}[]{c}
		1-a,-b+c,d,e\\
		1-a+b,c,-b+d,-b+e
	\end{array}
	\right]
	\, _3F_2
	\left( 
	\begin{array}[]{c}
		b,1+b-d,1+b-e\\
		1+b-a,1+b-c
	\end{array};1
	\right)
	\nonumber\\
	&+\Gamma
	\left[ 
	\begin{array}[]{c}
		1-a,b-c,d,e\\
		b,1-a+c,-c+d,-c+e
	\end{array}
	\right]
	\, _3F_2
	\left( 
	\begin{array}[]{c}
		c,1+c-d,1+c-e\\
		1-a+c,1-b+c
	\end{array};1
	\right)
	.
	\label{}
\end{align}
We additional apply the relation \cite{table}
\begin{align}
	_3F_2
	\left( 
	\begin{array}[]{c}
		a,b,c\\
		d,e
	\end{array};1
	\right)
	&=
	\Gamma
	\left[
	\begin{array}[]{c}
		d,d+e-a-b-c\\
		d+e-a-b,d-c
	\end{array}
	\right]
	\, _3F_2
	\left( 
	\begin{array}[]{c}
		e-a,e-b,c\\
		d+e-a-b,e
	\end{array};1
	\right)\,,
	\label{appB:tf1}
\end{align}
and obtain the following result
\begin{align}
	&
	\int_{C}^{}
	\frac{dz}{2\pi i}
	\Gamma
	\left[
	\begin{array}[]{c}
		a_1-z,a_2-z,b_1+z,b_2+z,b_3+z\\
		c+z
	\end{array}
	\right]
	\nonumber\\
	&\hspace{2cm}=
	\Gamma
	\left[
	\begin{array}[]{c}
		a_1+b_1,a_2+b_1,a_1+b_2,a_2+b_2,a_1+b_3,a_2+b_3\\
		a_{12}+{b_{13}},a_{12}+b_{23},-b_3+c
	\end{array}
	\right]
	\nonumber\\
	&\hspace{2cm}\quad \times
	\, _3F_2
	\left( 
	\begin{array}[]{c}
		a_1+b_3,a_2+b_3,a_{12}+b_{123}-c\\
		a_{12}+b_{13},a_{12}+b_{23}
	\end{array};1
	\right)
	\,.
	\label{appB:eb1}
\end{align}
This is the main result of this appendix. It reduces to the first Barnes
lemma for $c=b_j$ and to the second Barnes lemma for $c=a_{12}+b_{123}$.

The l.h.s. of Eq.~\eqref{appB:eb1} is symmetric in $a_1\leftrightarrow a_2$
and $b_1\leftrightarrow b_2 \leftrightarrow b_3$, however the symmetry among
$b_j$ is not obvious on the r.h.s.  We can show the symmetry by using the
transformation formula of $_3F_2$ given in Eq.~\eqref{appB:tf1}.

In general, the generalized hypergeometric function
$_3F_2(\{a,b,c\},\{d,e\};z)$ converges at $z=1$ when \cite{textbook}
\begin{align}
  \mathrm{Re}\, (d+e-a-b-c)>0\,.
\label{appB:conv}
\end{align}
This condition has to be satisfied when
using Eq.~\eqref{appB:eb1}.
If condition~\eqref{appB:conv} is violated we perform an analytic
continuation to obtain an expression which converges at $z=1$. 
This procedure is well-known \cite{textbook},
so we will not further discuss it here.

The convergence behaviour may change under the
replacements $b_1\leftrightarrow b_2 \leftrightarrow b_3$.
By applying the condition~\eqref{appB:conv} to $_3F_2$ of Eq.~\eqref{appB:eb1},
we obtain the condition $\mathrm{Re}\, (c-b_3)>0$,
which is clearly not symmetric under 
the replacements $b_1\leftrightarrow b_2 \leftrightarrow b_3$.
Thus the convergent domain, in terms of the space spanned by
$a_1,a_2,b_1,b_2,b_3$, can differ from expression to expression.

As an example, let us consider the integral
\begin{align}
\int _C
\frac{dz_1}{2\pi i}
\Gamma \left[ -z_1,-z_1-z_2,z_1,1+z_1+z_2 \right]
\psi (1+z_1)
\,.
\label{appB:example}
\end{align}
In this case, the right-most left-pole at $z_1=0$ merges with the left-most
right-pole. To separate the poles we introduce a regularization parameter
$\delta >0$ as $\Gamma (z_1)\to\Gamma (\delta +z_1)$ assuming
\begin{align}
-1<-\delta < \mathrm{Re}\, (z_1)< \mathrm{Re}\, (z_2)<0
\label{appB:cond}
\end{align}
and analytically continue $\delta \to 0$ later.  By applying
the replacements
\begin{align}
\{a_1\to 0\, ,
a_2\to -z_2\, ,
b_1\to 1\, ,
b_2\to \delta\, ,
b_3\to 1+z_2\, ,
c\to 1+\tilde c\,,
X\to 1+z_1\}
\end{align}
to Eq.~\eqref{appB:eb1}, we have
\begin{align}
	&
	\int_{C}^{}
	\frac{dz_1}{2\pi i}
	\Gamma
	\left[
	\begin{array}[]{c}
		-z_1,-z_1-z_2,1+z_1,\delta+z_1,1+z_1+z_2\\
		1+\tilde c+z_1
	\end{array}
	\right]
	\nonumber\\
	&\hspace{2cm}=
	\frac{1}{\delta -\tilde c}
	\left(
	\Gamma
	\left[
	\begin{array}[]{c}
		\delta, 1-z_2,z_2\\
		\tilde c
	\end{array}
	\right]
	-
	\Gamma
	\left[
	\begin{array}[]{c}
		\delta -z_2, 1-z_2,z_2\\
		\tilde c-z_2
	\end{array}
	\right]
	\right)
	\, ,
	\label{appB:eb2}
\end{align}
where the Gauss summation formula
\begin{align}
	\, _3F_2\left( 
	\begin{array}[]{c}
		a,b,d\\
		c,d
	\end{array};1
	\right)
	=
	\, _2F_1\left( 
	\begin{array}[]{c}
		a,b\\
		c
	\end{array};1
	\right)
	=
	\Gamma\left[ 
		\begin{array}[]{c}
			c-a-b,c\\
			c-a,c-b
		\end{array}
	\right]
	\,
	\label{}
\end{align}
has been used.  After differentiating w.r.t. $\tilde c$ and setting $\tilde
c\to 0$, we find
\begin{align}
	&
	\int_{C}^{}
	\frac{dz_1}{2\pi i}
	\Gamma
	\left[
	-z_1,-z_1-z_2,\delta+z_1,1+z_1+z_2
	\right]
	\psi (1+z_1)
	\nonumber\\
	&=
	-\frac{\gamma_E \Gamma [1+z_2,-z_2]}{\delta}
	+\frac{\Gamma [1+z_2,-z_2]}{2}
	\left(
	\gamma_E^2+\zeta_2+\psi(-z_2)^2-\psi '(-z_2)
	\right)
	+\mathcal{O}(\delta)
	\, .
	\label{appB:eb3}
\end{align}
Finally we can analytically continue $\delta \to 0$ as mentioned above; the r.h.s. becomes
\begin{align}
	&
	\int_{C}^{}
	\frac{dz_1}{2\pi i}
	\Gamma
	\left[
	-z_1,-z_1-z_2,\delta+z_1,1+z_1+z_2
	\right]
	\psi (1+z_1)
	\nonumber\\
	=&
	\int_{C}^{}
	\frac{dz_1}{2\pi i}
	\Gamma
	\left[
	-z_1,-z_1-z_2,z_1,1+z_1+z_2
	\right]
	\psi (1+z_1)
	\nonumber\\
	&-\frac{\gamma_E \Gamma [1+z_2,-z_2]}{\delta}
	+\Gamma [1+z_2,-z_2]
	\left(
	\gamma_E^2-\zeta_2-\gamma_E\psi(-z_2)+\gamma_E\psi(1+z_2)
	\right)
	+\mathcal{O}(\delta)
\end{align}
and therefore 
\begin{align}
	&
	\int_{C}^{}
	\frac{dz_1}{2\pi i}
	\Gamma
	\left[
	-z_1,-z_1-z_2,z_1,1+z_1+z_2
	\right]
	\psi (1+z_1)
	\nonumber\\
	&=
	\frac{\Gamma [1+z_2,-z_2]}{2}
	\left[
	-\gamma_E^2+\frac{\pi^2}{2}+2\gamma_E\psi(-z_2)
	+\psi(-z_2)^2-2\gamma_E\psi(1+z_2)-\psi'(-z_2)
	\right]
	\,.
\end{align}
Throughout this example, $z_2$ can assume any value satisfying
Eq.~\eqref{appB:cond}.  It can, in particular, be an integration variable.
Thus the two dimensional Mellin-Barnes integral of Eq.~(\ref{appB:example})
can be reduced to a one dimensional integral.


\end{appendix}



\begin{thebibliography}{99}
%
%

\bibitem{Glover:1987nx}
  E.~W.~N.~Glover and J.~J.~van der Bij,
  Nucl.\ Phys.\ B {\bf 309} (1988) 282.

\bibitem{Plehn:1996wb}
  T.~Plehn, M.~Spira and P.~M.~Zerwas,
  Nucl.\ Phys.\ B {\bf 479} (1996) 46
   Erratum: [Nucl.\ Phys.\ B {\bf 531} (1998) 655]
  [hep-ph/9603205].

\bibitem{Dawson:1998py}
  S.~Dawson, S.~Dittmaier and M.~Spira,
  Phys.\ Rev.\ D {\bf 58} (1998) 115012
  [hep-ph/9805244].

\bibitem{Grigo:2013rya}
  J.~Grigo, J.~Hoff, K.~Melnikov and M.~Steinhauser,
  Nucl.\ Phys.\ B {\bf 875} (2013) 1
  [arXiv:1305.7340 [hep-ph]].

\bibitem{Degrassi:2016vss}
  G.~Degrassi, P.~P.~Giardino and R.~Gröber,
  Eur.\ Phys.\ J.\ C {\bf 76} (2016) no.7,  411
  [arXiv:1603.00385 [hep-ph]].

\bibitem{Maltoni:2014eza}
  F.~Maltoni, E.~Vryonidou and M.~Zaro,
  JHEP {\bf 1411} (2014) 079
  [arXiv:1408.6542 [hep-ph]].

\bibitem{deFlorian:2013uza}
  D.~de Florian and J.~Mazzitelli,
  Phys.\ Lett.\ B {\bf 724} (2013) 306
  [arXiv:1305.5206 [hep-ph]].

\bibitem{deFlorian:2013jea}
  D.~de Florian and J.~Mazzitelli,
  Phys.\ Rev.\ Lett.\  {\bf 111} (2013) 201801
  [arXiv:1309.6594 [hep-ph]].

\bibitem{Grigo:2014jma}
  J.~Grigo, K.~Melnikov and M.~Steinhauser,
  Nucl.\ Phys.\ B {\bf 888} (2014) 17
  [arXiv:1408.2422 [hep-ph]].

\bibitem{Grigo:2015dia}
  J.~Grigo, J.~Hoff and M.~Steinhauser,
  Nucl.\ Phys.\ B {\bf 900} (2015) 412
  [arXiv:1508.00909 [hep-ph]].

\bibitem{Shao:2013bz}
  D.~Y.~Shao, C.~S.~Li, H.~T.~Li and J.~Wang,
  JHEP {\bf 1307} (2013) 169
  [arXiv:1301.1245 [hep-ph]].

\bibitem{deFlorian:2015moa}
  D.~de Florian and J.~Mazzitelli,
  JHEP {\bf 1509} (2015) 053
  [arXiv:1505.07122 [hep-ph]].

\bibitem{Ferrera:2016prr}
  G.~Ferrera and J.~Pires,
  JHEP {\bf 1702} (2017) 139
  [arXiv:1609.01691 [hep-ph]].

\bibitem{deFlorian:2016uhr}
  D.~de Florian, M.~Grazzini, C.~Hanga, S.~Kallweit, J.~M.~Lindert,
  P.~Maierhöfer, J.~Mazzitelli and D.~Rathlev,
  JHEP {\bf 1609} (2016) 151
  [arXiv:1606.09519 [hep-ph]].

\bibitem{Borowka:2016ypz}
  S.~Borowka, N.~Greiner, G.~Heinrich, S.~P.~Jones, M.~Kerner, J.~Schlenk and
  T.~Zirke,
  JHEP {\bf 1610} (2016) 107
  [arXiv:1608.04798 [hep-ph]].

\bibitem{Borowka:2016ehy}
  S.~Borowka, N.~Greiner, G.~Heinrich, S.~P.~Jones, M.~Kerner, J.~Schlenk,
  U.~Schubert and T.~Zirke,
  Phys.\ Rev.\ Lett.\  {\bf 117} (2016) no.1,  012001
   Erratum: [Phys.\ Rev.\ Lett.\  {\bf 117} (2016) no.7,  079901]
  [arXiv:1604.06447 [hep-ph]].

\bibitem{Heinrich:2017kxx}
  G.~Heinrich, S.~P.~Jones, M.~Kerner, G.~Luisoni and E.~Vryonidou,
  arXiv:1703.09252 [hep-ph].

\bibitem{Kudashkin:2017skd}
  K.~Kudashkin, K.~Melnikov and C.~Wever,
  arXiv:1712.06549 [hep-ph].

\bibitem{Bonciani:2016qxi}
  R.~Bonciani, V.~Del Duca, H.~Frellesvig, J.~M.~Henn, F.~Moriello and
  V.~A.~Smirnov,
  JHEP {\bf 1612} (2016) 096
  [arXiv:1609.06685 [hep-ph]].

\bibitem{Becchetti:2017abb}
  M.~Becchetti and R.~Bonciani,
  arXiv:1712.02537 [hep-ph].

\bibitem{Mastrolia:2017pfy}
  P.~Mastrolia, M.~Passera, A.~Primo and U.~Schubert,
  JHEP {\bf 1711} (2017) 198
  [arXiv:1709.07435 [hep-ph]].

\bibitem{Grober:2017uho}
  R.~Gr\"ober, A.~Maier and T.~Rauh,
  arXiv:1709.07799 [hep-ph].

\bibitem{progdata}
\verb|https://www.ttp.kit.edu/preprints/2018/ttp18-009/|.

\bibitem{Nogueira:1991ex}
  P.~Nogueira,
  J.\ Comput.\ Phys.\  {\bf 105} (1993) 279.

\bibitem{Harlander:1997zb}
  R.~Harlander, T.~Seidensticker and M.~Steinhauser,
  Phys.\ Lett.\ B {\bf 426} (1998) 125
  [hep-ph/9712228].

\bibitem{Seidensticker:1999bb}
  T.~Seidensticker,
  hep-ph/9905298.

\bibitem{Ruijl:2017dtg}
  B.~Ruijl, T.~Ueda and J.~Vermaseren,
  arXiv:1707.06453 [hep-ph].

\bibitem{Smirnov:2014hma}
  A.~V.~Smirnov,
  Comput.\ Phys.\ Commun.\  {\bf 189} (2015) 182
  [arXiv:1408.2372 [hep-ph]].

\bibitem{Lee:2012cn}
  R.~N.~Lee,
  arXiv:1212.2685 [hep-ph].

\bibitem{Lee:2013mka}
  R.~N.~Lee,
  J.\ Phys.\ Conf.\ Ser.\  {\bf 523} (2014) 012059
  [arXiv:1310.1145 [hep-ph]].

\bibitem{Kotikov:1990kg}
  A.~V.~Kotikov,
  Phys.\ Lett.\ B {\bf 254} (1991) 158.

\bibitem{Gehrmann:1999as}
  T.~Gehrmann and E.~Remiddi,
  Nucl.\ Phys.\ B {\bf 580} (2000) 485
  [hep-ph/9912329].

\bibitem{Melnikov:2017pgf}
  K.~Melnikov, L.~Tancredi and C.~Wever,
  Phys.\ Rev.\ D {\bf 95} (2017) no.5,  054012
  [arXiv:1702.00426 [hep-ph]].

\bibitem{Henn:2013pwa}
  J.~M.~Henn,
  Phys.\ Rev.\ Lett.\  {\bf 110} (2013) 251601
  [arXiv:1304.1806 [hep-th]].

\bibitem{Gituliar:2017vzm}
  O.~Gituliar and V.~Magerya,
  Comput.\ Phys.\ Commun.\  {\bf 219} (2017) 329
  [arXiv:1701.04269 [hep-ph]].

\bibitem{Meyer:2017joq}
  C.~Meyer,
  Comput.\ Phys.\ Commun.\  {\bf 222} (2018) 295
  [arXiv:1705.06252 [hep-ph]].

\bibitem{Beneke:1997zp}
  M.~Beneke and V.~A.~Smirnov,
  Nucl.\ Phys.\ B {\bf 522} (1998) 321
  [hep-ph/9711391].

\bibitem{Smirnov:2012gma}
  V.~A.~Smirnov,
  Springer Tracts Mod.\ Phys.\  {\bf 250} (2012) 1.

\bibitem{PSLQ}
H.R.P.~Ferguson and D.H.~Bailey, RNR Technical Report, RNR-91-032;
H.R.P.~Ferguson, D.H.~Bailey and S.~Arno, NASA Technical Report,
NAS-96-005.

\bibitem{Smirnov:2015mct}
  A.~V.~Smirnov,
  Comput.\ Phys.\ Commun.\  {\bf 204} (2016) 189
  [arXiv:1511.03614 [hep-ph]].

\bibitem{Pak:2010pt}
  A.~Pak and A.~Smirnov,
  Eur.\ Phys.\ J.\ C {\bf 71} (2011) 1626
  [arXiv:1011.4863 [hep-ph]].

\bibitem{Smirnov:1999wz}
  V.~A.~Smirnov and O.~L.~Veretin,
  Nucl.\ Phys.\ B {\bf 566} (2000) 469
  [hep-ph/9907385].

\bibitem{Bern:2005iz}
  Z.~Bern, L.~J.~Dixon and V.~A.~Smirnov,
  Phys.\ Rev.\ D {\bf 72} (2005) 085001
  [hep-th/0505205].

\bibitem{Czakon:2005rk}
  M.~Czakon,
  Comput.\ Phys.\ Commun.\  {\bf 175} (2006) 559
  [hep-ph/0511200].

\bibitem{barnesroutines}
  D. Kosower, \verb|https://mbtools.hepforge.org/|

\bibitem{Remiddi:1999ew}
  E.~Remiddi and J.~A.~M.~Vermaseren,
  Int.\ J.\ Mod.\ Phys.\ A {\bf 15} (2000) 725
  [hep-ph/9905237].

\bibitem{Borowka:2017idc}
  S.~Borowka, G.~Heinrich, S.~Jahn, S.~P.~Jones, M.~Kerner, J.~Schlenk and
  T.~Zirke,
  Comput.\ Phys.\ Commun.\  {\bf 222} (2018) 313
  [arXiv:1703.09692 [hep-ph]].

\bibitem{Anastasiou:2006hc}
  C.~Anastasiou, S.~Beerli, S.~Bucherer, A.~Daleo and Z.~Kunszt,
  JHEP {\bf 0701} (2007) 082
  [hep-ph/0611236].

\bibitem{table}
A.~P.~Prudnikov,
Yu.~A.~Brychkov,
O.~I.~Marichev
``Integrals and series, volume 3,''
Gordon and Breach Science Publishers (1990).

\bibitem{textbook}
L.~J.~Slater,
``Generalized Hypergeometric Functions,''
Cambridge University Press, (1966).

\end{thebibliography}
\end{document}